\documentclass[aps, twocolumn, a4paper, showpacs, superscriptaddress, floatfix, longbibliography]{revtex4-2} 
\usepackage[a4paper, centering, total={170mm,257mm}]{geometry}
\usepackage{amsmath, amssymb, physics}
\usepackage[hypertexnames=false]{hyperref}
\usepackage{datetime}
\usepackage{enumitem}
\usepackage{graphicx}
\usepackage{esdiff}
\usepackage{MnSymbol}
\usepackage{siunitx}
\usepackage{circuitikz}
\usepackage[normalem]{ulem}

\hypersetup{
    colorlinks,
    linkcolor={red!50!black},
    citecolor={blue!50!black},
    urlcolor={blue!80!black}
}

\renewcommand{\selectlanguage}[1]{} 

\def\vbr{\vb{r}}
\def\LS{\mathrm{LS}}
\def\wire{\mathrm{wire}}
\def\gate{\mathrm{gate}}

\footnotetext{\raggedright Correspondence should be addressed to mhuang@phys.ethz.ch or esslinger@ethz.ch}

\footnotetext{\raggedright These authors contributed equally to this work}

\begin{document}

\title{Irreversible entropy transport enhanced by fermionic superfluidity}

\author{Philipp Fabritius\textsuperscript{\dag}}
\author{Jeffrey Mohan\textsuperscript{\dag}}
\author{Mohsen Talebi}
\author{Simon Wili}
\affiliation{Institute for Quantum Electronics \& Quantum Center, ETH Zurich, 8093 Zurich, Switzerland}
\author{Wilhelm Zwerger}
\affiliation{Technische Universit\"at M\"unchen, Physik Department, 85748 Garching, Germany }
\author{Meng-Zi Huang\textsuperscript{*}}
\author{Tilman Esslinger\textsuperscript{*}}
\affiliation{Institute for Quantum Electronics \& Quantum Center, ETH Zurich, 8093 Zurich, Switzerland}

\date{22 April, 2024}

\begin{abstract}
The nature of particle and entropy flow between two superfluids is often understood in terms of reversible flow carried by an entropy-free, macroscopic wavefunction. While this wavefunction is responsible for many intriguing properties of superfluids and superconductors, its interplay with excitations in non-equilibrium situations is less understood. Here, we observe large concurrent flows of both particles and entropy through a ballistic channel connecting two strongly interacting fermionic superfluids. Both currents respond nonlinearly to chemical potential and temperature biases. We find that the entropy transported per particle is much larger than the prediction of superfluid hydrodynamics in the linear regime and largely independent of changes in the channel's geometry. In contrast, the timescales of advective and diffusive entropy transport vary significantly with the channel geometry. In our setting, superfluidity counterintuitively increases the speed of entropy transport. Moreover, we develop a phenomenological model describing the nonlinear dynamics within the framework of generalised gradient dynamics. Our approach for measuring entropy currents may help elucidate mechanisms of heat transfer in superfluids and superconducting devices. 
\end{abstract}

\maketitle

Two connected reservoirs exchanging particles and energy is a paradigmatic system that is key to understanding transport phenomena in diverse platforms of both fundamental and technological interest ranging from heat engines, to superconducting qubits \cite{senior_heat_2020}, and even heavy-ions collisions \cite{potel_quantum_2021}. Entropy and heat, both irreversibly produced and transported by the currents flowing between the reservoirs, are key quantities in superfluid and superconducting systems \cite{shelly_resolving_2016, fornieri_towards_2017}. They help to reveal microscopic information in strongly interacting systems \cite{ginzburg_thermoelectric_1989, crossno_observation_2016} and more generally characterise far-from-equilibrium systems \cite{ dogra_universal_2023}. Yet in traditional condensed matter systems such as superconductors and superfluid helium, the entropy is not directly accessible and requires indirect methods to deduce it \cite{bradley_direct_2011, pekola_colloquium_2021, ibabe_joule_2023}.

In this work, we leverage the advantage of quantum gases of ultracold atoms as naturally closed systems well-isolated from their environments to study entropy transport and production in fermionic superfluid systems. Using the known equation of state \cite{ku_revealing_2012}, we measure the particle number and total entropy in each of the two connected reservoirs as a function of evolution time, therefore directly obtaining the entropy current and production. In general, the nature of these currents depends on the coupling strength between the superfluids. On the one hand, two weakly coupled superfluids exhibiting the Josephson effect~\cite{valtolina_josephson_2015,luick_ideal_2020} exchange an entropy-free supercurrent described by Landau's hydrodynamic two-fluid model~\cite{patel_universal_2020,wang_hydrodynamic_2022,sidorenkov_second_2013,li_second_2022,yan_thermography_2024}. In quantum gases \cite{del_pace_tunneling_2021} as well as superconductors \cite{agrait_quantum_2003}, this is accomplished with low-transparency tunnel junctions weakly-biased in chemical potential or phase, while narrow channels are used to block viscous currents in superfluid helium \cite{hoskinson_transition_2006, botimer_pressure_2016}. On the other hand, superfluids strongly coupled by high-transparency channels \cite{krinner_observation_2014} can exhibit less intuitive behaviour since the supercurrent no longer dominates the normal current, making the system fundamentally non-equilibrium \cite{tinkham_introduction_1996, chen_dissipative_2014}. In particular, the signature of superfluidity in such systems is often large particle currents on the order of the superfluid gap which respond nonlinearly to chemical potential biases smaller than the gap. This is observed in ballistic junctions between superconductors~\cite{agrait_quantum_2003}, superfluid He \cite{viljas_multiple_2005}, and quantum gases~\cite{stadler_observing_2012, husmann_connecting_2015, huang_superfluid_2023}. However, entropy transport in this setting has so far only been experimentally studied indirectly and at higher temperatures in the linear response regime where the superfluidity of the system is ambiguous \cite{husmann_breakdown_2018, hausler_interaction-assisted_2021}, leaving open the question of entropy transport between strongly-coupled superfluids.

Here, we connect two superfluid unitary Fermi gases with a ballistic channel and measure the coupled transport of particles and entropy between them. We observe large sub-gap currents of both particles and entropy, indicating that the current is not a pure supercurrent and cannot be understood within a hydrodynamic two-fluid model. In particular, superfluidity counterintuitively enhances entropy transport in this system by enhancing particle current while maintaining a large entropy transported per particle. We also observe in our parameter regime that, while the system can always thermalise via the irreversible flow of this superfluid-enhanced normal current, thermalisation via pure entropy diffusion is inhibited in one-dimensional channels, giving rise to a non-equilibrium steady state previously observed in the normal phase \cite{husmann_breakdown_2018}. The observed nonlinear dynamics of particles and entropy are captured by a phenomenological model we develop whose only external constraints are the conservation of particles and energy and the Second Law of Thermodynamics.

\begin{figure}
    \includegraphics[width=\hsize]{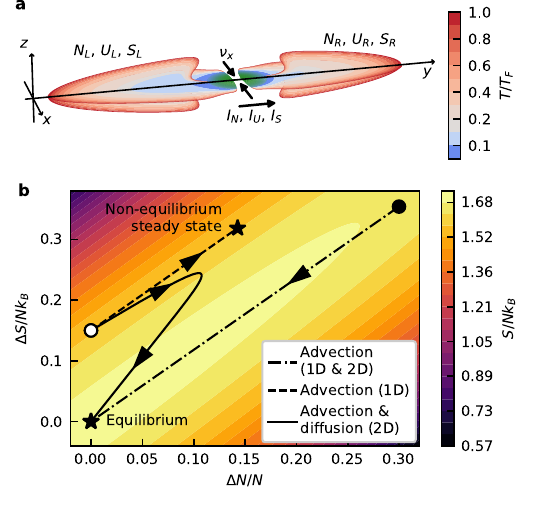}
    \caption{
    \textbf{Irreversible particle, energy, and entropy transport through a non-equilibrium channel connecting two superfluids.} \textbf{a}, Slices along the $x=0$ and $z=0$ planes of the calculated local degeneracy $T/T_F$ during transport through the 1D channel, showing that the reservoirs (left $L$ and right $R$) are in local equilibrium in the normal phase (red, $T/T_F>0.167$) over most of their volume and superfluid (blue, $T/T_F\leq0.167$) at the contacts to the channel. The channel (green) is far from local equilibrium. Differences in the atom number $N$, energy $U$, and entropy $S$ between the reservoirs induce currents between them. \textbf{b}, Depending on the initial conditions (filled or open circles) and microscopically allowed processes, the reservoirs can exchange entropy advectively and diffusively, tracing out various paths through state space with the constraints that atom number and energy are conserved $\dv*{N}{t}=\dv*{U}{t}=0$ and the entropy production is positive definite $\dv*{S}{t}\geq0$. This evolution halts (stars) by reaching either a non-equilibrium steady state ($\Delta N,\Delta S\neq0$) or equilibrium ($\Delta N=\Delta S=0$) where the total entropy $S=S_L+S_R$ has a global maximum.
    }
    \label{fig:fig1}
\end{figure}

We begin the experiment by preparing a balanced mixture of the first- and third-lowest hyperfine ground states of \textsuperscript{6}Li at unitarity in an augmented harmonic trap shown in Fig.~\ref{fig:fig1}a (Methods, Supplementary Section 3). To induce transport of atoms, energy, and entropy between the left ($L$) and right ($R$) reservoirs, we prepare an initial state within the state space in Fig.~\ref{fig:fig1}b characterised by the conserved total atom number and energy, $N=N_L+N_R$ and $U=U_L+U_R$, and the dynamical imbalances in the atom number $\Delta N=N_L-N_R$ and entropy $\Delta S=S_L-S_R$. The imbalances in the extensive quantities induce biases in the chemical potential $\Delta\mu=\mu_L-\mu_R$ and temperature $\Delta T=T_L-T_R$ according to the reservoirs' equations of state (EoS; see Methods) which in turn drive currents of the extensive properties $I_N(\Delta N, \Delta S) = -(1/2)\dv*{\Delta N}{t}$ and $I_S(\Delta N, \Delta S) = -(1/2)\dv*{\Delta S}{t}$. Note that $I_S$ is an apparent current, not a conserved current like $I_N$ and $I_U$, though we can place bounds on the conserved entropy current from the apparent current and entropy production rate $I_S^\mathrm{cons} \in [I_S-(1/2)\dv*{S}{t}, I_S+(1/2)\dv*{S}{t}]$ \cite{huang_superfluid_2023}. These equations of motion, together with the initial state $\Delta N(0), \Delta S(0)$, determine the path the system traces through state space $\Delta N(t), \Delta S(t)$ as well as the speed with which it traces this path. The paths that we explore experimentally are shown as black lines overlayed on top of the entropy landscape $S = S_L(N_L, U_L) + S_R(N_R, U_R) = S(N, U, \Delta N, \Delta S)$ computed from the EoS. The paths all exhibit a strictly positive entropy production rate $\dv*{S}{t}>0$, implying the transport is irreversible, until they reach either a non-equilibrium steady state ($\Delta N,\Delta S\neq0$) or equilibrium ($\Delta N=\Delta S=0$) where $S$ is maximised for the fixed $N$ and $U$. The von Neumann entropy of a closed system does not increase in time under Hamiltonian evolution, though the measured thermodynamic entropy $S$ can increase due to buildup of entanglement entropy shared between the two reservoirs \cite{breuer_theory_2002, esposito_entropy_2010, kaufman_quantum_2016, gnezdilov_ultrafast_2023}. We have verified that the system is nearly closed given the measured particle loss rate $\abs{\dv*{N}{t}}/N < \SI{0.01}{s^{-1}}$ and heating in equilibrium $\dv*{(S/N k_B)}{t} < \SI{0.02}{s^{-1}}$, limited by photon scattering of optical potentials, such that the entropy production observed during transport is due to the fundamental irreversibility of the transport.

We measure the system's evolution by repeatedly preparing the system in the same initial state, allowing transport for a time $t$, then taking absorption images of both spin states and extracting $N_i, S_i$ for both reservoirs $i=L,R$ using standard thermometry techniques (Methods, Supplementary Section 5B). Between the end of transport and imaging, we adiabatically ramp down the laser beams that define the channel to image the reservoirs in well-calibrated, half-harmonic traps. The beams do work on the reservoirs during this process and change $U_i$ but $N_i$ and $S_i$ remain constant. The cloud typically contains $N=270(30)\times10^3$ atoms and $S/N = 1.59(7) k_B$ before transport, below the critical value of $\approx1.90k_B$ for superfluidity in the transport trap. Figure~\ref{fig:fig1}a shows the local degeneracy $T/T_F$ in the $x=0$ and $z=0$ planes during transport calculated within the local density approximation \cite{haussmann_thermodynamics_2008} using the 3D equation of state \cite{ku_revealing_2012} to determine the local Fermi temperature $T_F(x,y,z)$ (Methods). The imbalance is illustrated using the representative values of $\nu_x=\SI{12.4}{kHz}$, $\Delta\mu = \SI{75}{nK} \times k_B$, and $\Delta T=0$. Assuming local equilibrium, the most degenerate regions in the channel would be deeply superfluid due to the strong, attractive gate potential and reach $T/T_F \approx 0.027$, $s \approx \num{7.2e-4}k_B$, and $\Delta/k_B T \approx 17$ where $s$ is the local entropy per particle and $\Delta$ is the superfluid gap. When $\nu_x\lesssim\SI{7}{kHz}$ (Methods, Supplementary Section 3) and the channel is 2D, normal regions appear at the edges of the channel that can also contribute to transport.

\begin{figure}
    \includegraphics{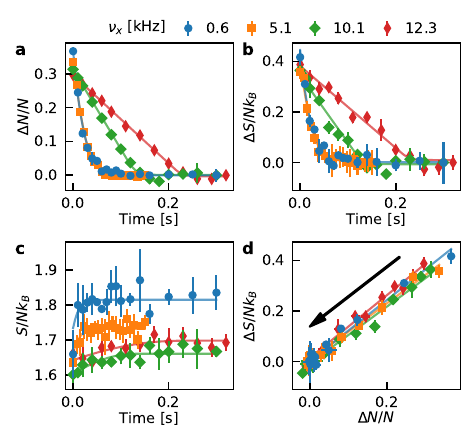}
    \caption{\textbf{Observing advective entropy current from the non-exponential evolution of imbalances in particle number and entropy}. Particles (\textbf{a}) and entropy (\textbf{b}) are transported between the reservoirs by currents that respond nonlinearly to biases in chemical potential and temperature, evidenced by the non-exponential decay. Both currents increase as the channel opens from 1D to 2D ($\nu_x$ decreasing). The net entropy (\textbf{c}) increases very slightly while $\Delta S$ changes significantly, indicating that entropy is indeed transported by the particle flow. Panel \textbf{d} shows $I_S=s^* I_N$ where the entropy advectively transported per particle $s^*$ is nearly independent of the channel geometry $\nu_x$ despite $I_N$ varying significantly. The observation that $s^*>0$ means the nonlinear particle current is not a superfluid current. Each data point is an average of 3-5 repetitions and error bars represent the standard deviation. Solid lines are fits of the phenomenological model.}
    \label{fig:fig2}
\end{figure}

In a first experiment, we prepare an initial state $\Delta N(0), \Delta S(0) \neq 0$ (filled circle in Fig.~\ref{fig:fig1}b) such that equilibrium is reached within $\SI{1}{s}$. For the strongest confinement, $\Delta N(t)$, shown in Fig.~\ref{fig:fig2}a, clearly deviates from exponential relaxation and the corresponding $I_N$ is much larger than the value $\sim \Delta\mu/h$ of a quantum point contact (QPC) in the normal state, indicating that the sub-gap current-bias characteristics are nonlinear (non-Ohmic) and the reservoirs are superfluid \cite{husmann_connecting_2015, huang_superfluid_2023, yao_controlled_2018}. When reducing $\nu_x$ to cross over from a 1D to 2D channel, $\Delta N(t)$ relaxes faster ($I_N$ increases) and, although it is less pronounced, the nonlinearity persists. The dynamics for the two smallest values of $\nu_x$ are nearly identical, suggesting that there are additional resistances in series with the 1D region such as the viscosity of the bulk reservoirs or the interfaces between the 3D and 2D regions \cite{hausler_interaction-assisted_2021} or between the normal and superfluid regions \cite{kanasz-nagy_anomalous_2016}.

Concurrently with $\Delta N(t)$, we observe non-exponential relaxation of $\Delta S(t)$ (Fig.~\ref{fig:fig2}b) which bears a remarkable resemblance to $\Delta N(t)$. We find that by plotting $\Delta S(t)$ against $\Delta N(t)$ in Fig.~\ref{fig:fig2}d, all paths collapse onto a single line. This demonstrates that the entropy current is directly proportional to the particle current $I_S = s^* I_N$ where the average entropy advectively transported per particle $s^*$ is nearly independent of $\nu_x$ even though $I_N$ itself varies significantly. Moreover, $\dv*{S}{t}$ (Fig.~\ref{fig:fig2}c) is barely resolvable and is significantly smaller than $I_S$, meaning there is indeed a large conserved entropy current flowing between the reservoirs. The dependence of $S/Nk_B$ on the confinement $\nu_x$ in Fig.~\ref{fig:fig2}c has a technical origin and does not affect the system during transport (Methods). The entropy transported per particle $s^* = 1.18(3) k_B$ is near its value in the normal phase \cite{husmann_breakdown_2018} and is orders of magnitude larger than the local entropy per particle in the channel assuming local equilibrium $s \approx \num{7.2e-4}k_B$. Because superfluidity in the contacts enhances $I_N$ while only slightly suppressing $s^*$, superfluidity increases $I_S$.

The fact that $s^*>0$ directly shows that the large, non-Ohmic current between the two superfluids is itself not a pure supercurrent in the context of a two-fluid model \cite{varoquaux_andersons_2015}. The observation that the flow is resistive is insufficient alone to conclude that it is not superfluid as there are many mechanisms for resistance to arise in a pure supercurrent \cite{likharev_superconducting_1979, halperin_resistance_2010}. The observation that $s^* \gg s$ suggests that the channel is far from equilibrium and hydrodynamics breaks down as is often the case in weak link geometries \cite{varoquaux_andersons_2015} unlike previous assumptions \cite{kanasz-nagy_anomalous_2016, hausler_interaction-assisted_2021}. We discuss in Supplementary Section 1 how the degree to which hydrodynamics breaks down depends on the preparation of the system. The large entropy current suggests an irreversible conversion process from superfluid currents in the contacts to normal currents in the channel and back to superfluid, or the propagation of normal currents originating in the normal regions of the reservoirs through the superfluids while remaining normal. Moreover, the independence of $s^*$ from $\nu_x$ implies that this process is independent of the channel geometry. There is an analogy between this observation and the central result of Landauer-B\"uttiker theory that the conductance through a ballistic channel is also independent of the geometry and depends only on the channel's transmission and the number of propagating modes.

\begin{figure}
    \includegraphics{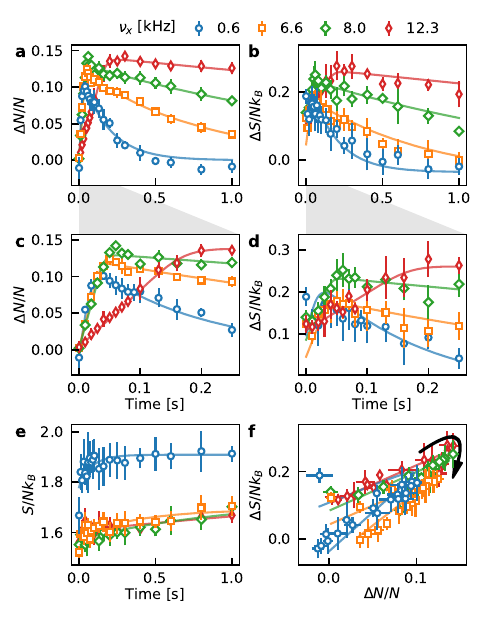}
    \caption{\textbf{Competition between two modes of entropy transport: nonlinear advection and linear diffusion.} A pure entropy imbalance induces currents of particles (\textbf{a}) and entropy (\textbf{b}) which are dominated by the nonlinear advective mode at early time and the linear diffusive mode at long times. For strong confinement (red) the system reaches a non-equilibrium steady state which persists beyond the experimentally accessible time, indicating the decoupling of advective and diffusive entropy transport. Reducing the confinement facilitates the diffusive mode, allowing equilibration and reducing the maximal response. \textbf{c},\textbf{d}, Zoomed-in views of the initial dynamics of \textbf{a},\textbf{b} for additional clarity. As in Fig.~\ref{fig:fig2}c, the increase in net entropy (\textbf{e}) indicates the irreversibility of the transport process. A fit with our phenomenological model (solid lines) describes the data well over the explored parameter regime. The competition between the two modes, which changes with $\nu_x$, determines the path traced through state space (\textbf{f}). Error bars (data points) indicate the standard deviation (average) of 3-5 repetitions.}
    \label{fig:fig3}
\end{figure}

In a second experiment, whose results are presented in Fig.~\ref{fig:fig3}, we prepare the system with a nearly pure entropy imbalance ($\Delta N(0)\approx0,\,\Delta S(0)\neq 0$, open circle in Fig.~\ref{fig:fig1}b). The initial response of $\Delta N$ and $\Delta S$ from $t=0$ to when $\Delta N$ reaches its maximum value is clearly non-exponential, resembling the advective dynamics in the first experiment with the same $s^*$, while the dynamics that follow are much slower and consistent with exponential relaxation and therefore linear response. With decreasing $\nu_x$, both dynamics become faster and the maximum values of $\Delta N$ and $\Delta S$ achieved at the turning point become smaller. For the largest value of $\nu_x$, the initial response is still fast while the relaxation that follows is extremely slow and resembles a non-equilibrium steady state over experimentally accessible timescales: the relaxation time of this state is $\SI{8(2)}{s}$ while it is reached from the initial state in only $\approx\SI{0.2}{s}$. In this state, the non-vanishing imbalances $\Delta N$ and $\Delta S$ depend on the initial state as well as the path of the system through state space determined by $s^*$, i.e., the system is non-ergodic. This indicates that the non-equilibrium steady state previously observed at higher temperatures \cite{husmann_breakdown_2018} persists in the superfluid regime where the current with which it is reached is $\gtrsim6$ times larger and non-Ohmic. Figure~\ref{fig:fig3}f shows the measured path in state space also illustrated in Fig.~\ref{fig:fig1}b. It demonstrates that the path is determined by the competition between the nonlinear and linear dynamics and varies with $\nu_x$, in contrast to the first experiment (Fig.~\ref{fig:fig2}d).

In the following, we formulate a minimal phenomenological model to describe our observations which are not captured by the linear response approach that successfully describes this system in the normal state \cite{husmann_breakdown_2018, hausler_interaction-assisted_2021} as it predicts purely exponential relaxation. We therefore turn to the formalism of generalised gradient dynamics \cite{pavelka_multiscale_2018}, a generalisation of Onsager's theory of irreversible processes (Methods, Supplementary Section 2). While it does not provide a microscopic theory, this formalism can describe general, irreversible, non-equilibrium processes and provides a convenient way to impose macroscopic constraints such as the Second Law of Thermodynamics and conservation laws for the particle number and energy. Within this framework, we make the Ansatz
\begin{align}\label{eq:pheno_model}
\begin{split}
    I_N &= I_\mathrm{exc} \tanh\pqty{\frac{\Delta\mu+\alpha_c\Delta T}{\sigma}} \\
    I_S &= \alpha_c I_N + G_T \Delta T/T
\end{split}
\end{align}
which produce entropy via the irreversible flow $\dv*{S}{t} = (I_N \Delta\mu + I_S \Delta T)/T$. The non-trivial result that $\alpha_c$ appears in both $I_N$ and $I_S$ is a generalisation of Onsager's reciprocal relations to nonlinear response and is a consequence of the irreversibility of these currents. The system exhibits two modes of entropy transport: a nonlinear advective mode $I_S^a = \alpha_c I_N$ characterised by the excess current $I_\mathrm{exc}$, Seebeck coefficient $\alpha_c$, and nonlinearity $\sigma$, wherein each transported particle carries entropy $s^*=\alpha_c$ on average, and a linear diffusive mode $I_S^d=G_T\Delta T/T$ characterised by the thermal conductance $G_T$ which enables entropy transport without net particle transport according to Fourier's law. The linear model is reproduced in the limit of large $\sigma$ with conductance $G=I_\mathrm{exc}/\sigma$. In a Fermi liquid, these two modes are related by the Wiedemann-Franz law where the Lorenz number $L = G_T/TG$ has the universal value $\pi^2 k_B^2/3$. The nonlinearity implies the breakdown of the Wiedemann-Franz law since the advective and diffusive modes are no longer linked \cite{husmann_breakdown_2018}. The excess current $I_\mathrm{exc}$ is the particle current with the nonlinearity saturated, as in superconducting weak links~\cite{likharev_superconducting_1979}.

\begin{figure}
    \includegraphics{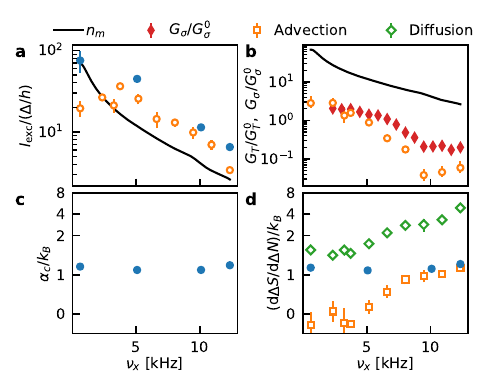}
    \caption{\textbf{Characterising the linear and nonlinear transport modes by fitting the phenomenological model across the 1D-2D crossover.} Fitted parameters from the first (second) experiment corresponding to Fig.~\ref{fig:fig2} (Fig.~\ref{fig:fig3}) are plotted with filled (open) markers here. \textbf{a}, The excess current normalised to the superfluid gap. \textbf{b}, The thermal and spin conductances normalised to the non-interacting values $G_T^0=2\pi^2k_B^2T/3h$, $G_\sigma^0=2/h$ for a single transverse mode (see main text). For reference, the number of occupied transverse modes in the channel $n_m$ is also shown in \textbf{a} and \textbf{b}. \textbf{c}, The Seebeck coefficient (advectively transported entropy per particle) extracted directly from the fits to the first experiment. \textbf{d}, The slope of the path through state space in the first experiment and in the advective ($\square$) and diffusive ($\diamond$) limits of the second experiment. Error bars represent standard errors from the fits.
    }
    \label{fig:fig4}
\end{figure}

Figure~\ref{fig:fig4} shows the parameters of the model as functions of $\nu_x$ extracted from the fits shown as curves in Figs.~\ref{fig:fig2} and \ref{fig:fig3}. Panel a shows $I_\mathrm{exc}$ normalised by the fermionic superfluid gap $\Delta/h$ along with the number of occupied transverse modes at equilibrium $n_m$ (Methods). Filled (open) circles were extracted from the first (second) experiment. $I_\mathrm{exc}$ follows $n_m\Delta/h$, increasing as $\nu_x$ decreases until $\nu_x\approx\SI{4}{kHz}$ where it plateaus, likely due to additional resistances in series with the 1D region. The fitted $I_\mathrm{exc}$ is apparently reduced in the second experiment relative to the first because the initial current is suppressed by the diffusive mode, making it more difficult to fit. It is intriguing that, consistent with previous studies \cite{husmann_connecting_2015, huang_superfluid_2023}, the equilibrium superfluid gap $\Delta$ within the local density approximation is still the relevant scale for the current, despite the evidence that the channel region is far from equilibrium.

Figure~\ref{fig:fig4}b shows $G_T$ and the spin conductance $G_\sigma$ separately measured in the same system by preparing a pure spin imbalance (Methods) normalised to their values for the single-mode non-interacting ballistic QPC $G_T^0 = 2\pi^2 k_B^2 T/3h,\, G_\sigma^0 = 2/h$. Both conductances are suppressed relative to the non-interacting values and increase monotonically with decreasing $\nu_x$, possibly due to the appearance of non-degenerate transverse modes at the edges of the channel (Methods, Supplementary Section 3). The non-equilibrium steady state arises from the fact that $G_T\rightarrow0$ in the 1D limit. The relative increase of $G_T$ with decreasing $\nu_x$ is larger than that of $G_\sigma$, suggesting that more types of excitations can contribute to diffusive entropy transport than spin transport, e.g., both collective phonon and quasiparticle excitations can contribute to $G_T$ \cite{uchino_role_2020} while only quasiparticle excitations can contribute to $G_\sigma$ \cite{sekino_mesoscopic_2020}. 

Figure~\ref{fig:fig4}c shows the fitted Seebeck coefficient $\alpha_c$ while Fig.~\ref{fig:fig4}d shows the slope of the path through state space $\dv*{\Delta S}{\Delta N}$ during the advective and diffusive dynamics. The fitted $\alpha_c$ and $\dv*{\Delta S}{\Delta N}$ match for the purely advective transport in the first experiment, showing that $\alpha_c$ is remarkably insensitive to $\nu_x$, while $\dv*{\Delta S}{\Delta N}$ more clearly shows how the two modes compete in the second experiment to determine the net response of the system. Figure~\ref{fig:fig4}d shows that, while both modes are generally present in the system's dynamics, their relative prevalence depends on $\nu_x$ as well as the initial state: the initial state in the first experiment was carefully chosen to allow the system to relax to equilibrium via the advective mode alone by preparing $\Delta S(0) = \alpha_c \Delta N(0)$ while the initial state in the second was chosen to contain both modes.

In summary, we have observed that the conceptually simple system of two superfluids connected by a ballistic channel exhibits the highly non-intuitive and currently unexplained effect that the presence of superfluidity increases the rate of irreversible entropy transport between them via nonlinear advection. This contrasts with the more familiar case of superfluid and superconducting tunnel junctions where the reversible, entropy-free Josephson current dominates. The entropy advectively carried per particle is nearly independent of the channel's geometry, while the timescales of advective and diffusive transport depends strongly thereon, raising the question of the microscopic origin of the observed entropy transported per particle $s^* \simeq 1 k_B$. Our phenomenological model that captures these observations, in particular the identification of advective $I_S^a \propto I_N$ and diffusive $I_S^d \propto \Delta T$ modes along with the sigmoidal shape of $I_N(\Delta\mu+\alpha_c\Delta T)$, may help guide future microscopic theories of the system. While extensive research has been conducted on the entropy producing effects of topological excitations of the superfluid order parameter~\cite{hoskinson_transition_2006, halperin_resistance_2010, bradley_direct_2011, silaev_universal_2012, barenghi_introduction_2014, varoquaux_andersons_2015, derrico_quantum_2017, valtolina_josephson_2015, burchianti_connecting_2018, wlazlowski_dissipation_2023}, less attention has been given to their influence on entropy transport and the possible pair-breaking processes they can induce. Early studies of superconductors found that mobile vortices can advectively transport entropy by carrying pockets of normal fluid~\cite{sensarma_vortices_2006} with them~\cite{solomon_thermomagnetic_1967, solomon_flux_1969, vidal_low-frequency_1973, tinkham_introduction_1996}. More generally, entropy-carrying topological excitations, which give rise to a finite chemical potential bias according to the Josephson relation $\Delta\mu = h \cdot \dv*{N_v}{t}$~\cite{varoquaux_andersons_2015}, can result from a complex spatial structure of the order parameter~\cite{wlazlowski_dissipation_2023}. Alternatively, it is possible that an extension of microscopic theories of multiple Andreev reflection~\cite{yao_controlled_2018, setiawan_analytic_2022}, which reproduce the finding that the excess current scales linearly with the number of channels and the gap~\cite{husmann_connecting_2015, huang_superfluid_2023}, may explain our observations. Clearly, a proper microscopic theory of this system is a challenge for the future. A complete understanding of the particle and entropy transport in superfluid systems is essential for both fundamental and technological purposes.

\section*{Methods} 
\textbf{Transport configuration.}
The atoms are trapped magnetically along $y$ and optically by a red-detuned beam along $x$ and $z$, with confinement frequencies $\nu_{\mathrm{trap},x} = \SI{171(1)}{Hz}$, $\nu_{\mathrm{trap},y} = \SI{28.31(2)}{Hz}$, and $\nu_{\mathrm{trap},z} =\SI{164(1)}{Hz}$. A pair of repulsive $\text{TEM}_{01}$-like beams propagating along $x$ and $z$, which we call the lightsheet (LS) and wire respectively, intersect at the centre of the trapped cloud and separate it into two reservoirs connected by a channel. The transverse confinement frequencies at the centre are set to $\nu_z=\SI{9.42(6)}{kHz}$ ($k_B T/h\nu_z = 0.21$) and $\nu_x=0.61\dots\SI{12.4(2)}{kHz}$ ($k_B T/h\nu_x=0.16\dots3.3$) with the powers of the beams. An attractive Gaussian beam propagating along $z$ with a similar size as the LS beam acts as a gate potential in the channel. The peak gate potential is $V_\mathrm{gate}^0 = -\SI{2.17(1)}{\micro K} \times k_B$. We use a wall beam, which is thin along $y$ and wide along $x$, during preparation and imaging to completely block transport with a barrier height $V_\mathrm{wall}^0$ much larger than $\mu$ and $k_B T$. The repulsive LS, wire, and wall are generated using blue-detuned \SI{532}{nm} light while the attractive gate is created with red-detuned \SI{766.7}{nm} light. 

The effective potential energy landscape along $y$ at $x=z=0$ (more details see Supplementary Section S3) is approximately $V_\mathrm{eff}(y,n_x,n_z)\approx h\nu_x(y)(n_x+1/2)+h\nu_z(y)(n_z+1/2)$, where $\nu_x$ and $\nu_z$ varies along $y$ due to the beams' profile and $n_x$, $n_z$ are the quantum numbers of the harmonic potential in $x$ and $z$ directions. The number of occupied transverse modes $n_m$ (Fig.~\ref{fig:fig4}) is calculated via the Fermi-Dirac occupation with local chemical potential set by $V_\mathrm{eff}(y,n_x,n_z)$
\begin{equation}
    n_m = \sum_{n_x, n_z = 0}^\infty \min_y \frac{1}{1 + \exp\qty{[V_\mathrm{eff}(y,n_x,n_z)-\mu]/k_B T}}\,,
\end{equation}
where the minimum occupation of each mode is used to account for modes that are not always occupied throughout the channel (Supplementary Section 3).

The complete potential energy landscape $V(\vb{r})$ was used to produce Fig.~\ref{fig:fig1}a via the local density approximation for the density $n(\vb{r}) = n[\mu-V(\vb{r}),T]$ \cite{ku_revealing_2012, haussmann_thermodynamics_2008} that determines the local Fermi temperature $k_B T_F(\vb{r}) = \hbar^2[3\pi^2n(\vb{r})]^{2/3}/2m$, where $m$ is the atomic mass.
The superfluid gap $\Delta$ assuming local equilibrium is estimated using the calculation in a homogeneous system $\Delta(\mu_c, T)$ \cite{haussmann_thermodynamics_2007} where $\mu_c = \max_{\vb{r}} [\mu - V(\vb{r})]$ is the maximum local chemical potential in the system. The crossover between 1D and 2D regimes ($\nu_x\sim\SI{7}{kHz}$) of the channel is estimated by comparing the local degeneracy along $x$ (at $y=z=0$) to the superfluid transition. In the 2D limit, non-superfluid modes can pass through the edges of the channel while in the 1D limit the degeneracy across the channel is below the superfluid transition (Supplementary Section 3). 

\textbf{Transport preparation.}
In order to prepare imbalances $\Delta S(0)$, $\Delta N(0)$ we ramp up the channel beams to separate the two reservoirs followed by forced optical evaporation. Using a magnetic field gradient along $y$, we shift the centre of the magnetic trap with respect to the channel beams before separation to prepare $\Delta N(0)$. By shifting the trap centre during evaporation, we can compress one reservoir and decompress the other, thereby changing their evaporation efficiencies and inducing a controllable $\Delta S(0)$. See Supplementary Section 4 for more details. To measure the spin conductance $G_\sigma$ (Fig.~\ref{fig:fig4}b), we prepare a ``magnetisation'' imbalance $\Delta M=\Delta N_\uparrow-\Delta N_\downarrow$. To do this, we ramp down the magnetic field before separating the reservoirs at \SI{52}{G} where the spins' magnetic moments are different and modulate a magnetic gradient along $y$ until the two spins oscillate out of phase. We then separate the two reservoirs and ramp back the magnetic field.

\textbf{Imaging and thermometry.}
Between the end of transport and the start of imaging, we ramp down the channel beams while keeping the wall on. At the end of each run, we obtain the column density $n_{i\sigma}^\mathrm{col}(y,z)$ of both reservoirs $i=L,R$ and both spin states $\sigma=\downarrow,\uparrow$ (first and third lowest states in the ground state manifold) from two absorption images taken in quick succession \textit{in situ}. We fit the degeneracy $q_{i\sigma} = \mu_{i\sigma}/k_B T_{i\sigma}$ and temperature $T_{i\sigma}$ of both reservoirs for each spin state using the EoS of the harmonically-trapped gas \cite{ku_revealing_2012}. However, we use the fitted temperature from the first image ($\downarrow$) for both spins since the density distribution in the second image is slightly perturbed by the first imaging pulse. The thermometry is calibrated using the critical $S/N$ of the condensation phase transition on the BEC side of the Feshbach resonance. See Supplementary Section 5B for more details.

\textbf{Generalized gradient dynamics.}
To ensure that the phenomenological nonlinear model satisfies basic properties such as the Second Law of Thermodynamics, it is formulated it in terms of a dissipation potential $\Xi$ \cite{pavelka_multiscale_2018}.
The thermodynamic fluxes are defined as derivatives of the dissipation potential $\Xi$ with respect to the forces $I_N = T\partial{\Xi}/\partial{\Delta\mu}$ and $I_S = T\partial \Xi/\partial\Delta T$. In this formalism, Onsager reciprocity and the conservation of particles and energy are fulfilled. Our model (Eq.~\ref{eq:pheno_model}) is the result of the following dissipation potential, which is constructed based on the experimental observation $I_S=s^* I_N$ and that $I_N$ follows a sigmoidal function of $\Delta \mu$~\cite{husmann_connecting_2015},
\begin{equation}
    \label{eq:superfluid_dissipation_potential_}
    \Xi = \frac{\sigma I_\mathrm{exc}}{T} \log\bqty{\cosh\pqty{\frac{\Delta\mu+s^*\Delta T}{\sigma}}}+\frac{G_T}{2}\pqty{\frac{\Delta T}{T}}^2\,,
\end{equation}
where the first part describes the advective and the second part the diffusive transport mode. See Supplementary Section 2 for more details.

\textbf{Reservoir thermodynamics.}
To formulate equations of motion in state space $(\Delta N, \Delta S)$, we relate $\Delta\mu, \Delta T$ to $\Delta N, \Delta S$ in terms of thermodynamic response functions
\begin{equation}
    \label{eq:reservoir_response_}
    \pmqty{\Delta N \\ \Delta S} \approx \frac{\kappa}{2} \pmqty{1 & \alpha_r \\ \alpha_r & \ell_r + \alpha_r^2} \pmqty{\Delta\mu \\ \Delta T}\,,
\end{equation}
where $\kappa$ is the compressibility, $\alpha_r$ is the dilatation coefficient, and $\ell_r$ is the ``Lorenz number'' of the reservoirs (Supplementary Section 5A, ref.~\cite{grenier_thermoelectric_2016}). 
To obtain the spin conductance, we use $\Delta M=(\chi/2)\Delta b$, where $\Delta b = (\Delta\mu_\uparrow-\Delta\mu_\downarrow)/2$. The spin susceptibility $\chi\approx0.32\kappa$, following the computed EoS of a polarised unitary Fermi gas~\cite{rammelmuller_finite-temperature_2018}.

The potential landscape $V(\vb{r})$ during the transport experiment deviates from simple harmonic potential due to the confinement beams as well as the anharmonicity of the optical dipole trap. We estimate from numeric simulations based on our knowledge of the $V(\vb{r})$ that $T$ 
and $\kappa$ agree within 1\% to those determined from absorption imaging in near-harmonic traps while $\mu$ is 24\% higher during transport. However, $\alpha_r$ and $\ell_r$ are more sensitive to the trap potential and can deviate by a factor of 3. We therefore fit these response coefficients in our model. See Supplementary Section 5A for more details. 

The total entropy $S$, being a state variable (contour plot in Fig.~\ref{fig:fig1}b), depends on the imbalances $\Delta N(t), \Delta S(t)$ but not the currents $I_N, I_S$. With the linearised reservoir response, the entropy produced by equilibration is given by
\begin{equation}
    S(t) \approx S_\mathrm{eq} - \frac{\Delta N^2(t)}{2 T \kappa} - \frac{[\Delta S(t) - \alpha_r \Delta N(t)]^2}{2 T \ell_r \kappa}\,,
    \label{eq:entropy_production}
\end{equation}
where $S_\mathrm{eq}$ is the maximum entropy at equilibrium given fixed total $N$ and $U$. The increase in $S/Nk_B$ with decreasing $\nu_x$ in Fig.~\ref{fig:fig2}c and Fig.~\ref{fig:fig3}c is caused by switching on the wall beam to block transport. For lower $\nu_x$, there are more atoms in the channel to be perturbed by this process.

\textbf{Fitting procedure.}
We fit the phenomenological model (Eq.~\ref{eq:pheno_model}) with linear reservoir responses (Eq.~\ref{eq:reservoir_response_}) to each data set--the set of different transport times at fixed $\nu_x$--independently for both the first and second experiment. We do this by solving the initial value problem for $\Delta N(t)$ and $\Delta S(t)$ given the parameters $\alpha_c$, $I_\mathrm{exc}$, $\sigma$, $G_T$ along with the reservoir response functions $\kappa$, $\alpha_r$, $\ell_r$. From these solutions, we also compute the total entropy $S(t)$ as a function of time (Eq.~\ref{eq:entropy_production}). We fit $\Delta N(t)$, $\Delta S(t)$ and $S(t)$ simultaneously to the data using a least-squares fit. For the first experiment, only $I_\mathrm{exc}$, $\sigma$, $\alpha_c$, $\Delta S(0)$ and $S_\mathrm{eq}$ are free parameters. Other parameters are fixed to their theoretical values for better fit stability. For the second experiment, we fit $\sigma$, $\Delta S(0)$, $S_\mathrm{eq}$, $G_T$, $\alpha_r$, $\ell_r$, and an offset in $\Delta S$ to account for drifts in alignment. $\alpha_c$ is fixed to the averaged value obtained in the first experiment. See Supplementary Section 6 for more details. The slopes shown in Fig.~\ref{fig:fig4}d are obtained by simple linear fits in the state space $\Delta S$ versus $\Delta N$ (Fig.~\ref{fig:fig2}d and Fig.~\ref{fig:fig3}f). The advective and diffusive modes in the second experiment are separated in time at the maximum $\Delta N(t)$ (cf.~Fig.~\ref{fig:fig2}a).

\section*{Acknowledgements}
We thank Alexander Frank for his contributions to the electronics of the experiment. We are grateful for inspiring discussions with the Madrid Quantum Transport Group (Alfredo Levy Yeyati, Gorm Steffensen, Francisco Matute-Ca\~nadas, Pablo Burset Atienza, Rafael S\'anchez), Panagiotis Christodoulou, Sarang Gopalakrishnan, Anne-Maria Visuri, Shun Uchino, Thierry Giamarchi, Alberto Montefusco, Boris Svistunov, and Selim Jochim. We thank Alex G\'omez Salvador and Eugene Demler for their productive and ongoing collaboration in investigating this system. We acknowledge the Swiss National Science Foundation (Grants No.~212168, UeM019-5.1, and TMAG-2\_209376) and European Research Council advanced grant TransQ (Grant No.~742579) for funding.

\bibliography{paper_sub}

\clearpage
\onecolumngrid
\section*{Supplemental Information}
\setcounter{equation}{0}
\setcounter{figure}{0}
\setcounter{section}{0}
\renewcommand{\thesection}{\arabic{section}}
\renewcommand{\thefigure}{S\arabic{figure}}
\renewcommand{\theequation}{S\arabic{equation}}

\def\vbr{\vb{r}}
\def\LS{\mathrm{LS}}
\def\wire{\mathrm{wire}}
\def\gate{\mathrm{gate}}

\section{Breakdown of hydrodynamics}
\label{sec:breakdown_of_hydrodynamics}

The two-fluid model states that the total particle current is the sum of the normal and superfluid (Josephson) components $I_N = I_N^n + I_N^s$ while the entropy current is carried advectively by only the normal current $I_S = s_n^* I_N^n$. We observe that the entropy current is proportional to the total particle current $I_S = s^* I_N$ which, when combined with the two previous equations, yields $I_N^s = (s_n^*/s^*-1) I_N^n$. The normal particle current $I_N^n$ is given by the normal particle current density $j_n$ integrated over the cross section at the centre of the channel and the entropy current is given by $j_n$ weighted by the local entropy per particle
\begin{equation}\begin{split}
    I_N^n &= \int\dd{x}\dd{z} j_n(x,y=0,z) \\
    I_S &= \int\dd{x}\dd{z} j_n(x,y=0,z) s(x,y=0,z).
\end{split}\end{equation}
If the system is hydrodynamic, the local entropy per particle is given by its value near equilibrium \cite{khalatnikov_introduction_2000, svistunov_superfluid_2015}, so the entropy transported per particle by the normal current would therefore be $s_n^* = I_S/I_N^n \sim s \approx \num{7.2e-4}k_B$. This is several orders of magnitude smaller than the observed entropy transported per particle $s^*=1.18(3)k_B$, implying that $I_N^s \approx - I_N^n$, i.e., the normal and superfluid currents flow in opposite directions and their small difference is the observed net current. Depending on whether or not the current-phase relation is multi-valued \cite{golubov_current-phase_2004}, the time-averaged supercurrent at finite $\Delta\mu$ either vanishes by undergoing reversible, adiabatic phase slips (AC Josephson effect) \cite{sols_crossover_1994, tinkham_introduction_1996} or flows in the same direction as the normal current by irreversibly nucleating topological excitations such as vortices~\cite{hoskinson_transition_2006}. A simple picture of this process is that the force $\propto-\Delta\mu$ accelerates the superfluid until it reaches its critical velocity, at which point it converts some kinetic energy to heat, then begins to accelerate again and the process repeats~\cite{varoquaux_andersons_2015}. This contradiction--that our observations imply $I_N^s\approx-I_N^n$ if the system is hydrodynamic while the equations of motion for the superfluid require $\mathrm{sign}(I_N^s)=\mathrm{sign}(I_N^n)$--shows that the hydrodynamic two-fluid model, namely that the transported entropy is determined by the near-equilibrium local entropy in the contacts, is not valid in the channel despite the internal equilibrium of the reservoirs sufficiently far from the channel (Sec.~\ref{sec:generalised_gradient_dynamics}). This argument does not rely on the precise values of $s$ and $s^*$, just on the fact that $s<s^*$. This means that the finding that hydrodynamic theory breaks down is robust to changes in $s^*$ over approximately three orders of magnitude.

As shown in the main text, superfluidity in the reservoirs increased the speed of entropy transport between them. In a bulk hydrodynamic system, entropy transport is faster in a superfluid, where entropy propagates as waves of second sound \cite{yan_thermography_2024}, than in a normal fluid where it propagates diffusively \cite{wang_hydrodynamic_2022}. Because it is a wave, second sound is in principle a reversible mechanism of particle and entropy transport, though the damping that often accompanies it \cite{li_second_2022} could allow it to transport entropy irreversibly as we observe here. Because the gas in the channel is far from equilibrium, second sound in a strict hydrodynamic sense \cite{pitaevskii_bose-einstein_2003} cannot propagate through it. Nevertheless, second sound in a more general, far-from-equilibrium sense referring to any relative motion between the normal and superfluid components induced by fundamental excitations \cite{donnelly_two-fluid_2009} may be present.

The channel and wall beams (Sec.~\ref{sec:potential_energy_landscape}) are on during the preparation of the system (Sec.~\ref{sec:preparation_of_the_initial_state}), while the gate beam is ramped on over \SI{10}{ms} followed by \SI{5}{ms} of settling before the wall is switched off to allow transport. 
We have verified that our observations of the advective mode, quantified by the values of $I_\mathrm{exc}$, $\sigma$, and $s^*$, remain unchanged for settling times up to \SI{0.5}{s} after the gate beam ramp time, much longer than the advective relaxation time (cf. Fig.~1). Longer settling times up to \SI{3.5}{s} lead to slower, more exponential relaxation of $\Delta N$ (smaller $I_\mathrm{exc}$ and larger $\sigma$), though the effect is consistent with heating induced by the gate beam. Nevertheless, using a high-resolution imaging system focused on the channel, we observe that the density in the region addressed by the gate previously referred to as pockets \cite{husmann_breakdown_2018} increases very slowly after the ramp on a timescale of several seconds, confirming the non-equilibrium nature of the channel.
The fact that the observed nonlinear dynamics are much faster than the local equilibration time of the channel and remain unchanged over a large range of preparation times suggests that the far-from-equilibrium state of the channel is robust and well-defined by the equilibrium reservoirs and channel geometry.

\section{Generalised gradient dynamics}
\label{sec:generalised_gradient_dynamics}

To ensure that our phenomenological nonlinear model satisfies some basic properties such as the Second Law of Thermodynamics and the conservation of particles and energy, it is useful to formulate it in terms of a dissipation potential $\Xi$ \cite{balian_microphysics_2007-1, pavelka_multiscale_2018}. Both Onsager's theory of irreversible processes \cite{landau_statistical_1969-1} and the more general formalism of gradient dynamics \cite{pavelka_multiscale_2018} identify thermodynamic forces and fluxes which determine the entropy production rate of a system $\dot{S} = \dv*{S}{t}$. This production rate $\dot{S}$ can be written in terms of thermodynamic observables from the first law of thermodynamics $\dd{S} = \sum_i x_i \dd{X}_i$ which relates the change in a system's entropy due to changes in the extensive quantities that characterise the system $X_i$ via their conjugate intensive variables $x_i$ \cite{landau_statistical_1969-1}, e.g., chemical potential, temperature, $s$-wave scattering length, normal and superfluid velocities, trap frequency, channel beam powers, etc. For a spin-balanced gas at unitarity in a static trap, this is
\begin{equation}
    \label{eq:first_law}
    \dd{S} = \frac{1}{T}\dd{U} - \frac{\mu}{T}\dd{N}.
\end{equation}
By dividing both sides of the equation by the time differential $\dd{t}$, we find the entropy production rate
\begin{equation}
    \label{eq:entropy_production_rate}
    \dot{S} = \dv{S}{t} = \frac{1}{T} \dv{U}{t} - \frac{\mu}{T} \dv{N}{t}
\end{equation}
which has the interpretation that if energy and atoms are added/removed at rates $\dot{U}$ and $\dot{N}$ while $\mu$ and $T$ are held constant, then the entropy increases/decreases at a rate $\dot{S}$ due to the thermalisation of non-equilibrium excitations produced by $\dot{U}$ and $\dot{N}$. Since the total entropy of the system is the sum of the entropies of the left and right reservoirs, the net entropy production rate
\begin{equation}
    \label{eq:net_entropy_production_rate}
    \dot{S} = \dot{S}_L + \dot{S}_R
\end{equation}
is given in terms of the production rates in the two reservoirs
\begin{equation}
    \label{eq:reservoir_entropy_production_rate}
    \dot{S}_i = \frac{1}{T_i} \dot{U}_i - \frac{\mu_i}{T_i} \dot{N}_i.
\end{equation}
Because the total energy and atom number is conserved
\begin{equation}\begin{split}
    \label{eq:conservation_laws}
    \dot{U} &= \dot{U}_L + \dot{U}_R = 0 \\
    \dot{N} &= \dot{N}_L + \dot{N}_R = 0,
\end{split}\end{equation}
their addition/removal rates in each reservoir are given by the conserved currents flowing between them
\begin{equation}\begin{split}
    I_U &= -\frac{1}{2}(\dot{U}_L - \dot{U}_R) = -\frac{1}{2}\dv{\Delta U}{t} \\
    I_N &= -\frac{1}{2}(\dot{N}_L - \dot{N}_R) = -\frac{1}{2}\dv{\Delta N}{t}
\end{split}\end{equation}
Combining this with Eq.~\ref{eq:net_entropy_production_rate} and \ref{eq:reservoir_entropy_production_rate}, we find that $\dot{S}$ can be written equivalently in terms of either $I_N$ and $I_U$ or, without any approximations,  $I_N$ and the apparent entropy current $I_S=-(1/2)\dv*{\Delta S}{t}$
\begin{equation}
    \dot{S} = I_N \Delta\pqty{\frac{\mu}{T}} + I_U \Delta\pqty{-\frac{1}{T}}
    = \frac{I_N \Delta\mu + I_S \Delta T}{T}
\end{equation}
where $T=(T_L+T_R)/2$. This means we can formulate a theory with either the fluxes $I_N$ and $I_U$ driven by the forces $\Delta(\mu/T)=\mu_L/T_L-\mu_R/T_R$ and $\Delta(-1/T)=-1/T_L+1/T_R$ or the fluxes $I_N$ and $I_S$ driven by the forces $\Delta\mu/T$ and $\Delta T/T$ or simply $\Delta\mu$ and $\Delta T$. We chose the latter as the entropy is more natural to measure given our experimental sequence (the energy is not conserved by the channel beam ramps but entropy is) and because it gives deeper insight into the nature of the flow between the two reservoirs. The vector of state variables is therefore $\vb{X} = (N, U, \Delta N, \Delta S)$ and the vector of their conjugate quantities are $\vb{x} = \pdv*{S}{\vb{X}} = (-\mu/T, 1/T, -\Delta\mu/2T, -\Delta T/2T)$, though practically they can be reduced to $\vb{X} = (\Delta N, \Delta S)$ and $\vb{x} = (-\Delta\mu/2T, -\Delta T/2T)$ because $\dot{N}=\dot{U}=0$. The energy and entropy currents are related by
\begin{equation}
    I_U = \pqty{T - \frac{\Delta T^2}{4T}} I_S + \pqty{\mu - \frac{\Delta \mu \Delta T}{4T}} I_N
\end{equation}
so, for small biases $\Delta T \ll T, \Delta\mu \ll \mu$, the entropy current is approximately given by the heat current
\begin{equation}
    \label{eq:heat_current}
    I_Q = I_U - \mu I_N \approx T I_S.
\end{equation}

The fluxes are defined as derivatives of the dissipation potential $\Xi$ with respect to the forces
\begin{equation}
\label{eq:currents_from_dissipation_potential}
    I_N = \pdv{\Xi}{\Delta(\mu/T)},~I_U = \pdv{\Xi}{\Delta(-1/T)} \quad\text{or}\quad I_N = T\pdv{\Xi}{\Delta\mu},~ I_S = T\pdv{\Xi}{\Delta T} 
\end{equation}
Since $\Xi$ is a single-valued scalar potential function of two scalars, Onsager reciprocity
\begin{equation}
    \pdv{I_N}{\Delta(-1/T)} = \pdv{I_U}{\Delta(\mu/T)}
    \qq{or}
    \pdv{I_N}{\Delta T} = \pdv{I_S}{\Delta\mu}
\end{equation}
is naturally fulfilled in this formalism at any $\Delta(\mu/T), \Delta(-1/T)$ or $\Delta\mu,\Delta T$, not just in the linear response regime near equilibrium as is usually the case when imposing this condition.

To formulate equations of motion for the evolution in state space $\Delta N(t), \Delta S(t)$, we need to relate $\Delta\mu, \Delta T$ to $\Delta N, \Delta S$ via the reservoirs' equation of state (EoS). For the relatively small imbalances used here, we can linearise the EoS in terms of thermodynamic response functions
\begin{equation}
    \label{eq:reservoir_response}
    \pmqty{\Delta N \\ \Delta S} \approx \frac{\kappa}{2} \pmqty{1 & \alpha_r \\ \alpha_r & \ell_r + \alpha_r^2} \pmqty{\Delta\mu \\ \Delta T}
\end{equation}
where $\kappa$ is the compressibility, $\alpha_r$ is the dilatation coefficient, and $\ell_r$ is the ``Lorenz number'' of the reservoirs \cite{grenier_thermoelectric_2016}; see Sec.~\ref{sec:reservoir_thermodynamics} for definitions. Within the same approximation, the energy and entropy imbalance are related by the average chemical potential $\mu=(\mu_L+\mu_R)/2$ and temperature $T=(T_L+T_R)/2$
\begin{equation}
    \label{eq:Delta_U}
    \Delta U \approx \mu \Delta N + T \Delta S.
\end{equation}

This formalism obeys the principle that $S$ is a state variable: $S$ depends only on $N$, $U$, $\Delta N$, and $\Delta S$ but not on the path the system travelled in state space. In other words, the entropy produced by equilibration is the entropy difference between the initial and final states
\begin{equation}\begin{split}
    \delta S &= \int_0^\infty\dd{t} \dot{S}(t)
    = \int_0^\infty\dd{t} \frac{I_N(t) \Delta\mu(t) + I_S(t) \Delta T(t)}{T(t)} \\
    &= \int_0^\infty\dd{t} \dot{\vb{X}} \cdot \vb{x}
    = \int_{\vb{X}_0}^{\vb{X}_\infty}\dd{\vb{X}} \cdot \pdv{\Xi}{\vb{X}} \\
    &= S(N,U,0,0) - S(N,U,\Delta N_0, \Delta S_0)
\end{split}\end{equation}
and is independent of the microscopic processes and current-bias characteristics of the system contained in $\Xi$. If we use the linearised reservoir response in Eq.~\ref{eq:reservoir_response}, this becomes
\begin{equation}
    \label{eq:entropy_produced}
    \delta S \approx \frac{\Delta N_0^2}{2 T \kappa} + \frac{(\Delta S_0 - \alpha_r \Delta N_0)^2}{2 T \ell_r \kappa}
\end{equation}
which, together with $S(N,U,0,0)$, is the expression plotted in Fig.~1b. The interpretation of this result is that there is initially some potential energy in the system due to the imbalances, like the charging energy of a capacitor, which, due to the Joule heating from irreversible flow through the resistive channel, is converted into heat energy, i.e., entropy. We have verified that the system is nearly closed by ensuring the absence of particle loss $\abs{\dv*{N}{t}}/N < \SI{0.01}{s^{-1}}$ and heating in equilibrium $\dv*{(S/N k_B)}{t} < \SI{0.02}{s^{-1}}$ due to, e.g., photon scattering from the beams, vacuum background collisions, and three-body recombination \cite{chin_feshbach_2010}. The total duration of the sequence is kept constant, varying only the amount of time the channel is open, so the entropy production plotted in Fig.~2c and Fig.~3e includes the small and constant offset coming from residual heating not due to irreversible transport. The increase in $S/Nk_B$ with decreasing $\nu_x$ in Fig.~2c and Fig.~3e is caused by switching on the wall beam to block transport. This process affects low $\nu_x$ more since there are many more atoms in the channel in that case to absorb the potential energy. This additional heat does not affect the state of the system during transport as it is only added afterwards.

This formalism with the state vector $\vb{X} = (N, U, \Delta N, \Delta S)$ only applies when the reservoirs are in internal equilibrium, i.e., the time evolution is quasi-stationary. If, for example, they were hydrodynamic and their breathing modes were excited, then the amplitudes and phases of these modes would have to be added to the state vector $\vb{X}$ to completely characterise the state of the system. Once the complete set of experimentally distinguishable degrees of freedom are added to $\vb{X}$, this formalism applies again even away from equilibrium. The fact that our observations are consistent with a description using only these four state variables of the reservoirs confirms the assumptions that the atom number and energy are conserved and that the reservoirs are in internal equilibrium during transport. It is therefore natural to ask whether this formalism still applies when the channel is far from equilibrium and in principle contains many degrees of freedom that must be added to $\vb{X}$. Indeed it does because the state of the channel is fixed by its boundary conditions imposed by the equilibrium reservoirs. Therefore, each additional degree of freedom of the channel is a function of only the four state variables in $\vb{X}$ and need not be added for a complete description.

Generalised gradient dynamics can only account for irreversible dynamics, however a phase bias $\Delta\phi$ between the superfluids can in principle drive a reversible Josephson supercurrent $I_N \sim I_c \sin(\Delta\phi)$ which transports and produces no entropy \cite{zhao_heat_2004}. Our observations show that the dynamics are irreversible in nature since the particle current both transports and produces entropy and we observe no undamped oscillations characteristic of the Josephson effect in a finite system \cite{valtolina_josephson_2015, luick_ideal_2020}. It is nevertheless somewhat surprising that we do not observe this effect, however this is likely a consequence of the high transmission of the channel. In a ballistic quantum point contact (QPC), the critical supercurrent $I_c^\mathrm{QPC} = \Delta/\hbar$ \cite{martin-rodero_microscopic_1994} and the excess normal current $I_\mathrm{exc}^\mathrm{QPC} = 16\Delta/3h$ \cite{cuevas_hamiltonian_1996} are comparable. This normal current can flow at small bias $\Delta\mu/\Delta$ in ballistic channels but not tunnel junctions \cite{meier_josephson_2001, cuevas_hamiltonian_1996} and can therefore shunt the supercurrent and damp reversible oscillations \cite{yao_controlled_2018}. If both irreversible and reversible dynamics are significant, then they can both be incorporated into a phenomenological theory in the form of a general equation for non-equilibrium reversible-irreversible coupling (GENERIC) \cite{ottinger_beyond_2005, pavelka_multiscale_2018}.

In the normal system where linear response theory is valid \cite{husmann_breakdown_2018, hausler_interaction-assisted_2021}, the dissipation potential and currents it generates are
\begin{equation}\begin{split}
    \label{eq:linear_model}
    \Xi_n &= \frac{G}{2T} (\Delta\mu + \alpha_c\Delta T)^2 + \frac{G_T}{2}\pqty{\frac{\Delta T}{T}}^2
    = \frac{G}{2T} [(\Delta\mu + \alpha_c\Delta T)^2 + L \Delta T^2]
    = \Xi_n^a + \Xi_n^d \\
    \pmqty{I_N \\ I_S} &= G\pmqty{1 & \alpha_c \\ \alpha_c & L + \alpha_c^2} \pmqty{\Delta\mu \\ \Delta T}.
\end{split}\end{equation}
This model exhibits a linear advective mode of entropy transport $I_S^a = \alpha_c G(\Delta\mu + \alpha_c \Delta T) = \alpha_c I_N$ characterised by the conductance $G$ and Seebeck coefficient $\alpha_c$, and a linear diffusive mode $I_S^d = G_T \Delta T/T$ characterised by the thermal conductance $G_T$ or equivalently the Lorenz number $L = G_T/T G$.

In the superfluid system, the experimental observation $I_S = s^* I_N$ simplifies the problem of identifying $\Xi$ since it implies the existence of an advective mode $\Xi_s^a$. Eq.~\ref{eq:currents_from_dissipation_potential} imposes that $\Xi_s^a$ must satisfy $(\pdv*{\Xi_s^a}{\Delta T})_{\Delta \mu} = s^* (\pdv*{\Xi_s^a}{\Delta\mu})_{\Delta T}$, which has the unique solution $\Xi_s^a(\Delta\mu, \Delta T) = \Xi_s^a(\Delta\mu + s^* \Delta T)$. Based on our observations that $I_N$ is a sigmoidal function of $\Delta\mu$ \cite{husmann_connecting_2015, huang_superfluid_2023}, we make the Ansatz
\begin{equation}
    \label{eq:superfluid_dissipation_potential}
    \Xi_s^a = \frac{\sigma I_\mathrm{exc}}{T} \log\bqty{\cosh\pqty{\frac{\Delta\mu+s^*\Delta T}{\sigma}}}\,.
\end{equation}
In fact, we find that the simplest form that describes our observations from both the first and second experiment is the sum of the nonlinear advective and linear diffusive modes $\Xi = \Xi_s^a + \Xi_n^d$, which generates the currents
\begin{align}\begin{split}
    \label{eq:currents_full_form}
    I_N &= I_\mathrm{exc} \tanh\pqty{\frac{\Delta\mu+\alpha_c\Delta T}{\sigma}} \\
    I_S &= \alpha_c I_N + G_T \Delta T/T
\end{split}\end{align}
where the Seebeck coefficient $\alpha_c$ takes the place of the entropy advectively transported per particle $s^*$. It is readily verifiable that $\Xi_n^a$, $\Xi_n^d$, and $\Xi_s^a$ satisfy the six formal criteria of dissipation potentials \cite{pavelka_multiscale_2018}.  Eq.~\ref{eq:currents_full_form} shows that entropy transport in the superfluid system has the same form as the normal system but the advective mode now responds nonlinearly to the thermodynamic force $\Delta\mu+\alpha_c\Delta T$ and is characterised by the excess current $I_\mathrm{exc}$ \cite{huang_superfluid_2023}, originally defined as the additional current above the normal, Ohmic current in superconducting tunnel junctions \cite{agrait_quantum_2003}, and the sharpness of the nonlinearity $\sigma$. Ballistic QPCs of superconductors and superfluid \textsuperscript{3}He exhibit similar current-bias characteristics where the nonlinearity is determined by the highest order multiple Andreev reflection process the QPC can support, i.e., the maximum number of Cooper pairs that can be transported in a single coherent process $\sigma\sim\Delta/n_\mathrm{pair}$ \cite{agrait_quantum_2003, viljas_multiple_2005}. The nonlinearity implies the breakdown of the Wiedemann-Franz law since the advective and diffusive modes are no longer linked. The linear model (Eq.~\ref{eq:linear_model}) is reproduced in the limit of large $\sigma$ with $G=I_\mathrm{exc}/\sigma$. Our procedure to fit this model to the data is described in Sec.~\ref{sec:fitting_procedure}. Our fits are also consistent with the coexistence of both linear and nonlinear advective modes $\Xi = \Xi_s^a + \Xi_n^a + \Xi_n^d$ so we cannot exclude a finite linear advective mode $\Xi_n^a$ but find that the nonlinear advective mode $\Xi_s^a$ adequately describes our observations, yielding reduced chi-square statistics slightly below 1.

\section{Potential energy landscape}
\label{sec:potential_energy_landscape}
The total potential energy landscape is a combination of the trap, the channel beams, and the spatially-varying zero point energy of the channel beams' confinement $V(\vbr) = V_\mathrm{trap}(\vbr) + V_\mathrm{ch}(\vbr) + V_\mathrm{ZPE}(y)$. Our trap is a combination of a Gaussian optical dipole trap with wavelength $\lambda=\SI{1064}{nm}$ and a magnetic field oriented along $z$ with harmonic curvature, giving a combined trap potential
\begin{equation}\begin{split}
    V_\mathrm{trap}(\vbr) &= V_\mathrm{mag}(\vbr) + V_\mathrm{dip}(\vbr) \\
    V_\mathrm{mag}(\vbr) &= \frac{1}{2}m(2\pi\nu_{\mathrm{mag},y})^2(x^2 + y^2 - 2z^2) \\
    V_\mathrm{dip}(\vbr) &= V_\mathrm{dip}^0\qty{1 - \frac{w_{\mathrm{dip},x,0} w_{\mathrm{dip},z,0}}{w_{\mathrm{dip},x}(y) w_{\mathrm{dip},z}(y)} \exp\bqty{-2\frac{x^2}{w_{\mathrm{dip},x}^2(y)}-2\frac{z^2}{w_{\mathrm{dip},z}^2(y)}}}     
\end{split}\end{equation}
where $w_{\mathrm{dip},x/z}(y)$ follows the usual hyperbolic Gaussian beam width divergence, $\nu_{\mathrm{mag},y} = \SI{28.30(2)}{Hz}$, $w_{\mathrm{dip},x}(0) = w_{\mathrm{dip},z}(0) = \SI{80.7(1)}{\micro m}$, $V_\mathrm{dip}^0 = \SI{1.33(1)}{\micro K} \times k_B$. When $\mu, k_B T \ll V_\mathrm{dip}^0$, the trap is approximately harmonic with confinement frequencies in all three spatial directions $\nu_{\mathrm{trap},x} = \SI{171(1)}{Hz}$, $\nu_{\mathrm{trap},y} = \SI{28.31(2)}{Hz}$, $\nu_{\mathrm{trap},z} =\SI{164(1)}{Hz}$ and the average trap frequency $\bar{\nu}_\mathrm{trap} = \SI{92.7(5)}{Hz}$.

A pair of repulsive $\text{TEM}_{01}$-like beams propagating along $x$ and $z$, which we call the lightsheet (LS) and wire respectively, intersecting at the centre of the trapped cloud separate it into two reservoirs connected by a channel. The transverse confinement frequencies at their centre are $\nu_z=\SI{9.42(6)}{kHz}$ ($k_B T/h\nu_z = 0.21$) and $\nu_x$ ranges from \SI{0.61(1)}{kHz} to \SI{12.4(2)}{kHz} ($k_B T/h\nu_x \in [0.16, 3.3]$) depending on the power of the beam such that we can explore the crossover from one dimension (1D) to two dimensions (2D) in the channel. An attractive Gaussian beam propagating along $z$ acts as a gate potential in the channel. The combined potential energy landscape of these three beams is
\begin{equation}\begin{split}
    V_\mathrm{ch}(\vbr) &= V_\LS(\vbr) + V_\wire(\vbr) + V_\gate(\vbr) \\
    V_\LS(\vbr) &= f_\LS(y) \qty{V_\LS^0 + \frac{\pi^3 m (\nu_{\LS,z} w_{\LS,z})^2}{2} \bqty{\mathrm{erfi}(z/w_{\LS,z}) }^2f_\LS(z) } \\
    V_\wire(\vbr) &= f_\wire(y) \qty{V_\wire^0 + m(\pi \nu_{\wire,x} w_\wire^\mathrm{notch})^2 f_\wire(x)\bqty{1 - f_\wire^\mathrm{notch}(x)}} \\
    V_\gate(\vbr) &= f_\gate(y) f_\gate(x) V_\gate^0 
\end{split}\end{equation}
The envelope function for each beam $f_b(y) \propto P_b e^{-2(y/w_{b,y})^2}$ determine their spatial profile along the respective direction and are proportional to the power of each beam $P_b$. The finite potential of the LS and wire at the origin $V_{\LS/\wire}^0$ arise from optical aberrations that causes their nodal planes to not be perfectly dark, but this is a small correction on the order of 0.1\% of the peak potential of each beam. The repulsive LS and wire are generated using blue-detuned \SI{532}{nm} light while the attractive gate is created with red-detuned \SI{766.7}{nm} light. The beam waists are $w_{\LS,y} = \SI{30.2}{\micro m}$, $w_{\LS,z} = \SI{9.5}{\micro m}$, $w_{\wire,x} = \SI{78}{\micro m}$, $w_{\wire,y} = \SI{6.82}{\micro m}$, $w_\wire^\mathrm{notch} = \SI{1.5}{\micro m}$, $w_{\gate,x} = \SI{30.4}{\micro m}$, $w_{\gate,y} = \SI{31.8}{\micro m}$, and the peak confinement frequencies are $\nu_{\LS,z} = \SI{9.42(6)}{kHz}$ and $\nu_{\wire,x} = \SI{12.4(2)}{kHz}$ at the reference beam powers used for calibration where $f_{\LS/\wire}(0)=1$. The peak gate potential is $V_\gate^0 = -\SI{2.17(1)}{\micro K} \times k_B$.

The confinement of the channel beams along $x$ and $z$ raise the zero point energy (ZPE) of the atoms, effectively producing a new potential energy along $y$ relative to the ZPE of the trap
\begin{equation}
    V_\mathrm{ZPE}(y) = \frac{h}{2} \left(\nu_x(y) +  \nu_z(y) - (\nu_{\mathrm{trap},x} + \nu_{\mathrm{trap},z})\right) 
\end{equation}
with the transversal confinement $\nu_x(y) = \sqrt{f_\wire(y) \nu_{\wire,x}^2 + f_\gate(y) \nu_{\gate,x}^2 + \nu_{\mathrm{trap},x}^2}$ and $ \nu_z(y) = \sqrt{f_\LS(y) \nu_{\LS,z}^2 + \nu_{\mathrm{trap},z}^2}$ and the confinement frequency of the gate beam given by $\nu_{\gate,x} = \sqrt{-V_\gate^0/4\pi^2 m w_{\gate,x}^2}$. The higher lying modes defined by the transverse confinement that can contribute to transport have energies determined by the quantum numbers of the transverse harmonic oscillator states $n_{x/z} = 0, 1, \dots$ Their effective potential energy landscape above the trap potential is
\begin{equation}
    V_\mathrm{eff}(y,n_x,n_z) = V_\mathrm{ZPE}(y) + V_\mathrm{ch}(0,y,0) + h\nu_x(y) n_x + h\nu_z(y) n_z
\end{equation}
which determines number of occupied modes that contribute to transport
\begin{equation}
    n_m = \sum_{n_x, n_z = 0}^\infty \min_y \frac{1}{1 + \exp\qty{[V_\mathrm{eff}(y,n_x,n_z)-\mu]/k_B T}}.
\end{equation}
The minimum occupation of each mode is used to account for modes that may be occupied in the centre of the channel but are unoccupied elsewhere and therefore non-propagating \cite{hausler_interaction-assisted_2021}. The particle conductance of a non-interacting gas through a contact with perfect transparency is $G=2n_m/h$ \cite{krinner_observation_2014}.

The wall beam, which is only on during preparation and imaging and not during transport, is similar to the wire beam but without the notch cut out
\begin{equation}
    V_\mathrm{wall}(\vbr) = f_\mathrm{wall}(y) V_\mathrm{wall}^0 e^{-2(x/w_{\mathrm{wall},x})^2}
\end{equation}
where the barrier height $V_\mathrm{wall}^0$ is larger than $\mu$ and $k_B T$ to completely block transport. Its width is large enough to completely suppress tunnelling \cite{kwon_strongly_2020}.

\begin{figure}
    \centering
    \includegraphics{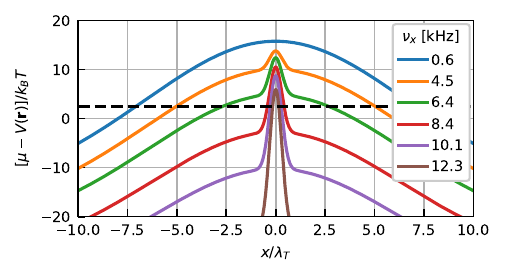}
    \caption{\textbf{1D-2D crossover of the channel.} The equilibrium local degeneracy at $y=z=0$ vs. $x$ in units of the thermal de Broglie wavelength $\lambda_T=h/\sqrt{2\pi mk_BT}=\SI{2.4}{\micro m}$ for various powers of the wire beam shows that for high powers (large $\nu_x$), the channel is tightly confining and there are few occupied modes (regions with degeneracy $>0$) while for low powers (small $\nu_x$), the confinement is very weak and there are many occupied modes. Since, in the 2D limit, the potential varies over a much larger length scale than $\lambda_T$, we can apply a local density approximation picture and identify regions at large $x$ (the edges of the channel) which are occupied and can contribute to transport but below the critical degeneracy to be superfluid (horizontal dashed line).}
    \label{fig:1d2d}
\end{figure}

The complete potential energy landscape $V(\vbr)$ was used to produce Fig.~1a via the local density approximation for the density $n(\vbr) = n[\mu-V(\vbr),T]$ \cite{ku_revealing_2012, haussmann_thermodynamics_2008} that determines the local Fermi temperature $k_B T_F(\vbr) = \hbar^2[3\pi^2n(\vbr)]^{2/3}/2m$. For weak wire beam powers, the transverse confinement frequency at the centre of the channel $\nu_x=\nu_x(y=0)$ is small and the barriers at the edges of the cloud disappear. In this case, the equilibrium degeneracy varies slowly along $x$ from deeply degenerate and superfluid in the centre of the channel to weakly degenerate and normal at the edges of the channel to vacuum far from the centre. Fig.~\ref{fig:1d2d}a shows how the degeneracy varies with $x$ (in units of the thermal de Broglie wavelength $\lambda_T = h/\sqrt{2\pi m k_B T} = \SI{2.4}{\micro m}$) around $\vbr=0$ for various powers of the wire beam. The horizontal dashed line is the critical degeneracy at the superfluid transition $(\mu/k_B T)_c = 2.49$ \cite{ku_revealing_2012}. This shows that in the 1D channel, the only occupied modes are tightly confined and expected to be superfluid at equilibrium, while in the 2D channel, weakly confined and normal transport modes appear at the edges. While the location of the 1D-2D transition is not well-defined, the channel is significantly wider at the occupation threshold $\mu-V(\vb{r})=0$ for $\nu_x\lesssim\SI{7}{kHz}$. This is approximately the same $\nu_x$ as the centre location of the feature resembling an avoided crossing in Fig.~4d, which is determined by the dimensional crossover. The 2D contacts to the 1D channel are the most degenerate regions in the system in equilibrium, see Fig.~1a. We estimate the superfluid gap here using the local density approximation applied to a calculation of the gap in a homogeneous system $\Delta(\mu_c, T)$ \cite{haussmann_thermodynamics_2007} where $\mu_c = \max_{\vbr} [\mu - V(\vbr)]$.

\section{Preparation of the initial state}
\label{sec:preparation_of_the_initial_state}

As part of the experimental sequence, we ramp up the channel beams to separate the two reservoirs then performed forced optical evaporation before allowing transport by opening the wall. Using a magnetic field gradient along $y$, we can shift the centre of the magnetic trap with respect to the channel beams before separation to prepare $\Delta N(0)$ before shifting the centre back to coincide with the channel beams for transport, which ensures that $\Delta N=0$ at equilibrium $\Delta\mu=0$. By shifting the centre after separation to a different position during evaporation, we can compress one reservoir and decompress the other, thereby changing their evaporation efficiencies and inducing a controllable $\Delta S(0)$ as well. The average efficiency changes only slightly with the imbalances such that the net entropy per particle $S/N k_B$ varies by $<9\%$ over the large range of $\Delta N(0)$ and $\Delta S(0)$ we prepare. This method achieves much lower entropy--therewith enabling the study of entropy transport in the superfluid phase--than a previously used method \cite{brantut_thermoelectric_2013, husmann_breakdown_2018, hausler_interaction-assisted_2021} where evaporation was performed in the same trap as separation, but additional entropy was injected into one reservoir by focusing the gate beam into the reservoir and modulating the power.

\begin{figure}
    \centering
    \includegraphics{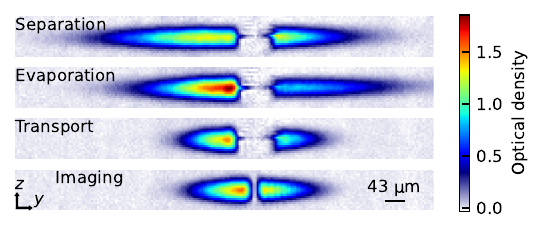}
    \caption{\textbf{Absorption images of the system at various stages of the experiment}: just after separation into two reservoirs, during evaporation, during transport without the gate beam, and during imaging. Each image is an average over 10 shots of spin $\ket{\uparrow}$ and the magnified pixel size is \SI{4.3}{\micro m}.}
    \label{fig:preparation}
\end{figure}

Between the end of transport and the start of imaging, we ramp down the channel beams while keeping the wall on. Typical images of the cloud at each of these stages are shown in Fig.~\ref{fig:preparation}. For the experiments at all $\nu_x$ that contain only the advective modes (Fig.~2), we prepare a density imbalance $\Delta N(0)$ and the corresponding entropy imbalance $\Delta S(0)$ such that the system relaxes to equilibrium in the 1D channel $\nu_x=\SI{12.4(2)}{kHz}$ at long times and not the NESS. For the experiments with both the advective and diffusive modes (Fig.~3), we fix $\Delta N(0)=0$ and prepare as large $\Delta S(0)$ as possible without losing atoms.

To measure the spin conductance $G_\sigma$, we employ a similar method as in \cite{krinner_mapping_2016} though instead of quenching the gradient along $y$, we modulate it at the average trap frequency for the two spin states for half the period of the trap frequency difference to induce a larger initial spin imbalance. We use the recently computed equation of state of the finite-temperature polarised unitary fermi gas \cite{rammelmuller_finite-temperature_2018} to compute a spin susceptibility $\chi \approx 0.32\kappa$ \cite{huang_superfluid_2023} in agreement with measurements on a harmonically-trapped unitary fermi gas at low temperatures \cite{sommer_universal_2011}.

\section{Reservoir thermodynamics}
\label{sec:reservoir_thermodynamics}

\subsection{Effect of the potential energy landscape}
\label{subsec:effect_of_the_potential_energy_landscape}

The potential energy landscape $V(\vbr)$ described in Sec.~\ref{sec:potential_energy_landscape} determines the equilibrium thermodynamic properties of the reservoirs. In the local density approximation \cite{haussmann_thermodynamics_2008}, the grand canonical free energy of the full system (both spins in both reservoirs) is given in terms of the universal pressure equation of state of the unitary fermi gas \cite{ku_revealing_2012}
\begin{equation}
    \Omega(\mu,T) = -\frac{k_B T}{\lambda_T^3} \int\dd[3]{r} f_P\bqty{\frac{\mu-V(\vbr)}{k_B T}}\,.
\end{equation}
While $\Omega$ takes a simple form for a harmonic trap, the trap is highly anharmonic during transport due to the channel beams (especially the lightsheet) and is slightly anharmonic during imaging due to the wall and gaussian dipole trap. It is therefore a non-universal function of $\mu$, $T$, and the powers of the various beam $\Omega(\mu,T,P_\mathrm{dip},P_\mathrm{LS},P_\mathrm{wire},P_\mathrm{gate},P_\mathrm{wall})$. The magnetic trap frequency $\nu_{y,\mathrm{mag}}$ is fixed at unitarity by the geometry of the Feshbach coils. In a given trap defined by these powers, the extensive quantities and response functions are given by derivatives of $\Omega$
\begin{equation}\begin{split}
    N &= -\pqty{\pdv{\Omega}{\mu}}_T \\
    S &= -\pqty{\pdv{\Omega}{T}}_\mu \\
    U &= \Omega + \mu N + TS \\
\end{split}
\quad\quad
\begin{split}
    \kappa &= -\pdv[2]{\Omega}{\mu} 
    \\ \alpha_r &= -\frac{1}{\kappa}\pdv[2]{\Omega}{\mu}{T} 
    \\ \ell_r &= -\frac{1}{\kappa}\pdv[2]{\Omega}{T} - \alpha_r^2 
\end{split}
\end{equation}

As explained in Sec.~\ref{sec:potential_energy_landscape} and shown in Fig.~\ref{fig:preparation}, the potential energy landscape $V(\vbr)$ differs between transport, where we wish to know the response coefficients, and imaging, where we measure $N_i, S_i$, so the EoS of the gas $\Omega(\mu,T)$ is different in these two conditions. However, the two are connected because $N_i, S_i$ are the same in both the transport and imaging conditions as a consequence of the adiabaticity of the channel beam ramps. This fact can be used to determine the chemical potential and temperature in each reservoir during transport $\mu_i^\mathrm{tr}, T_i^\mathrm{tr}$ from their values extracted in the imaging condition $\mu_i^\mathrm{im}, T_i^\mathrm{im}$ using the EoS of each reservoir in the two configurations $\Omega^\mathrm{tr}(\mu_i^\mathrm{tr}, T_i^\mathrm{tr})/2$ and $\Omega^\mathrm{im}(\mu_i^\mathrm{im}, T_i^\mathrm{im})/2$ that can be computed from the known $V^\mathrm{tr}(\vbr)$ and $V^\mathrm{im}(\vbr)$. The response coefficients can then be computed from $\mu^\mathrm{tr} = (\mu_L^\mathrm{tr} + \mu_R^\mathrm{tr})/2$, $T^\mathrm{tr} = (T_L^\mathrm{tr} + T_R^\mathrm{tr})/2$, and $\Omega^\mathrm{tr}(\mu^\mathrm{tr}, T^\mathrm{tr})$.

The lightsheet has the strongest influence on the reservoirs' EoS since it is large in comparison to the cloud as seen in Fig.~\ref{fig:preparation}. Its primary effect is to increases the chemical potential by pushing the cloud to the edges of the trap where the potential energy is larger. We observe this directly by imaging the cloud in the transport configuration for various lightsheet beam powers and see that, while the atom number in the system remains constant, atoms are pushed away from the central region. The gate has a weaker but still significant effect at the strong lightsheets used here: it draws about 7\% of the total number of atoms into the channel independently of $\nu_x$. The local density approximation with the 3D EoS $f_P$ overestimates the fraction of atoms drawn into the channel since the system is in fact quasi-2D or quasi-1D in most of the channel where the density is lower at the same $\mu$ and $T$ \cite{fenech_thermodynamics_2016, boettcher_equation_2016}, though this has a small effect on the overall thermodynamics as the 2D and 1D regions are small. Using the procedure outlined above, we estimate the ratio of several thermodynamic properties during transport to their values during imaging. We estimate that $T^\mathrm{tr}$ is within 1\% of $T^\mathrm{im}$ while $\mu^\mathrm{tr}$ is 24\% larger than $\mu^\mathrm{im}$. Furthermore, $\kappa^\mathrm{tr}$ is within 1\% of $\kappa^\mathrm{im}$ but $\alpha_r^\mathrm{tr}$ and $\ell_r^\mathrm{tr}$ are respectively 3.4 and 2.6 times larger than their values during imaging. In short, $T$ and $\kappa$ are essentially the same during transport as during imaging, with the other parameters, especially the response functions, are much more sensitive to the trap potential. This is consistent with our observation that the conductance of the weakly-interacting gas is quantised in units of $1/h$ which relies on an accurate value of $\kappa$ without incorporating these corrections \cite{krinner_observation_2014}. Motivated by this finding and the fact that the potential energy landscape we estimate is idealised and not exact (there are clear asymmetries in the lightsheet in Fig.~\ref{fig:preparation}), we fix $T$ and $\kappa$ to the measured values but fit $\alpha_r$ and $\ell_r$ along with the other parameters of the model (Sec.~\ref{sec:fitting_procedure}).

\subsection{Thermometry from absorption images}
\label{subsec:thermometry_from_absorption_images}

It is not straightforward to apply standard thermometry techniques \cite{ketterle_making_2008} to absorption images of the reservoirs in the transport configuration including all the channel beams (see Fig.~\ref{fig:preparation}) since it is difficult to precisely characterise the complex potential energy landscape that determines the reservoirs' EoS. We therefore ramp the channel beams down between the end of transport and the beginning of imaging such that the potential energy landscape is well-known and harmonic. This ramp is performed adiabatically such that the atom number and entropy in each reservoir remains constant even though their energies change due to the work done by the time-varying beams. This was confirmed by ensuring that $N_i$ and $S_i/N_i k_B$ are the same for slower ramps.

At the end of each shot, we obtain the column density $n_{i\sigma}^\mathrm{col}(y,z)$ of both reservoirs $i=L,R$ and both spin states $\sigma=\downarrow,\uparrow=\ket{1},\ket{3}$ from two absorption images \SI{225}{\micro s} apart taken \textit{in situ} along the $x$ axis with a calibrated imaging system. The atom number in each spin state and reservoir is determined by integrating the column density over the half-plane of each reservoir
\begin{equation}\begin{split}
    N_{L\sigma} &= \int_{-\infty}^0 \dd{y} \int_{-\infty}^\infty \dd{z} n_{L\sigma}^\mathrm{col}(y,z) \\
    N_{R\sigma} &= \int_0^\infty \dd{y} \int_{-\infty}^\infty \dd{z} n_{R\sigma}^\mathrm{col}(y,z).
\end{split}\end{equation}
The centre $y=0$ is fixed to the centre of a gaussian fit to the density profile in the wall region. We fit the degeneracy $q_{i\sigma} = \mu_{i\sigma}/k_B T_{i\sigma}$ and temperature $T_{i\sigma}$ of both reservoirs for each spin state using the EoS of the harmonically-trapped gas
\begin{equation}
    n_{i\sigma}^\mathrm{col}(y,z) = \lambda_{Ti\sigma}^{-3} \int_{-\infty}^\infty \dd{x} f_n\bqty{q_{i\sigma} - \pqty{\frac{x}{R_{xi\sigma}}}^2 - \pqty{\frac{y}{R_{yi\sigma}}}^2 - \pqty{\frac{z-z_0}{R_{zi\sigma}}}^2}
\end{equation}
where $\lambda_{T i\sigma} = h/\sqrt{2\pi m k_B T_{i\sigma}}$ is the thermal de Broglie wavelength, $f_n$ is the universal density EoS of a single spin in a balanced unitary Fermi gas \cite{ku_revealing_2012}, and $R_{ji\sigma}^2 = 2k_B T_{i\sigma}/m(2\pi\nu_{\mathrm{trap},j})^2$ is the Gaussian thermal length given by the known trap frequencies and the fitted temperature. In practice, we do not fix $R_z$ to $R_y$ via the trap frequencies and the common temperature, but rely only on the calibrated and almost harmonic magnetic trap frequency along $y$. We exclude the region $\approx\SI{60}{\micro m}$ wide containing the wall from the fit, fix both reservoirs and spin states to have a common centre $y=0$ fixed to the fitted centre of the wall, and fit $z_0$ with a common value for the two reservoirs. We also fit a small rotation angle between the camera's CCD grid and the harmonic trap principle directions, which we find to be $0.45(5)^\circ$. We estimate that, at the same $\mu$ and $T$, the gas with the wall of finite width has $S/N k_B$ that is $<2\%$ larger than the half-harmonic approximation wherein the wall is assumed to have zero width. While the atom numbers match for the two spin states, we see that the fitted $q$ for the second spin state to be imaged (regardless of which spin is imaged first) is systematically larger, likely due to off-resonant photon scattering from the first pulse and energy transfer from collisions with the the first spin. We therefore use the thermodynamics of the first spin state. This thermometry method based on directly fitting $q_{i\sigma}$ and $T_{i\sigma}$ gives similar results as a previously-used method using the second spatial moment of the reservoirs' column density distributions to extract thermodynamic properties \cite{husmann_breakdown_2018}.

\subsection{Calibration of absorption images}
\label{subsec:calibration_of_absorption_images}

The absorption images of the column density are determined from raw image containing atoms $A(y,z)$ and a ``bright'' reference image $B(y,z)$
\begin{equation}
    \label{eq:OD_from_images}
    \frac{\sigma_0}{\alpha} n_\mathrm{col}(y,z) = \mathrm{OD}(y,z) = \log\frac{B(y,z)}{A(y,z)} + \frac{B(y,z) - A(y,z)}{\chi t}
\end{equation}
where $\sigma_0$ is the resonant absorption cross section, $\alpha$ is a correction of the absorption cross section, $\chi$ is the imaging transition's saturation intensity $I_\mathrm{sat}$ in units of the camera count rate, and $t$ is the pulse duration \cite{horikoshi_appropriate_2017, hueck_calibrating_2017}. For each shot, we use the optimal reference image $B$ \cite{ockeloen_detection_2010} to suppress photon shot noise and technical noise, though the latter contribution is negligible. The duration of the imaging pulses was $t=\SI{2}{\micro s}$ to avoid any Doppler shifts, which we verified by ensuring that the resonance frequency does not depend on the pulse power. We calibrated $\chi$ by ensuring the measured atom number does not depend on the imaging intensity $I$ \cite{reinaudi_strong_2007}; we ultimately used $I/I_\mathrm{sat}\approx0.75$.

\begin{figure}
    \centering
    \includegraphics{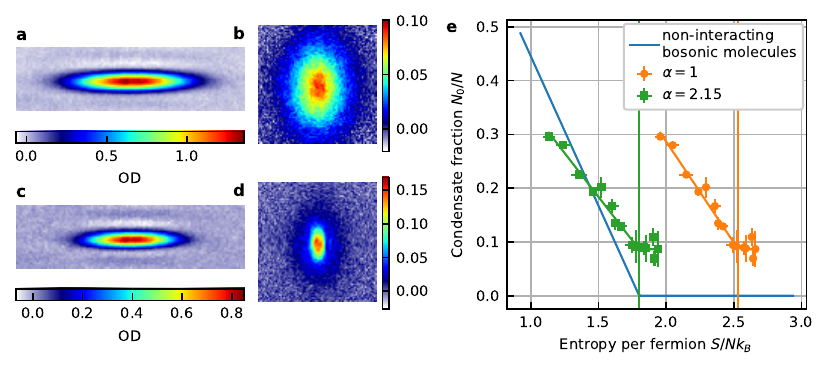}
    \caption{\textbf{Calibration of the absorption images via Bose-Einstein condensation of molecules.} \textbf{a}, Average of 5 absorption images of a single spin acquired \textit{in situ} in the harmonic trap at unitarity with a relatively deep dipole trap at the end of the forced optical evaporation to prepare a non-degenerate cloud. \textbf{b}, The same system prepared in \textbf{a} after an adiabatic sweep of the magnetic field to the BEC side of the Feshbach resonance followed by a short time of flight. The density distribution is well-fitted by a gaussian and is therefore above the critical point for Bose-Einstein condensation. \textbf{c}, Same as \textbf{a} but with a lower dipole trap depth at the end of evaporation to prepare a more degenerate cloud. \textbf{d}, Same as \textbf{c} after the magnetic field sweep and time of flight, displaying the bimodal density distribution that indicates Bose-Einstein condensation of Feshbach molecules. \textbf{e}, The condensate fraction measured on the BEC side vs. the entropy per fermion measured at unitarity. The solid line shows the expected behaviour for a non-interacting Bose gas, the orange circles show the measurement with $\alpha=1$, and the green squares show the measurement when correcting the absorption cross section with $\alpha=2.15$ to match the measured and theoretical critical points.}
    \label{fig:condensate_fraction_vs_entropy_per_particle}
\end{figure}

The correction $\alpha$ is typically calibrated by imaging a very degenerate cloud and fixing the amplitude of the OD using the equation of state \cite{ku_thermodynamics_2015, horikoshi_appropriate_2017}, however the clouds that we prepare are not sufficiently degenerate to robustly fit $q$ and $\alpha$ independently. We therefore calibrated $\alpha$ by measuring the onset of Bose-Einstein condensation after an adiabatic sweep of the magnetic field to the BEC side of the Feshbach resonance \cite{ketterle_making_2008} as shown in Fig.~\ref{fig:condensate_fraction_vs_entropy_per_particle}. We prepared harmonically trapped gases (no channel beams) at varying levels of degeneracy by varying the depth of the optical dipole trap at the end of evaporation and then either imaged \textit{in situ} to perform the thermometry procedure described in \ref{subsec:thermometry_from_absorption_images} or adiabatically swept the magnetic field over \SI{50}{ms} to the BEC regime and imaged the cloud after a \SI{6}{ms} time of flight. The magnetic field at the end of the sweep is \SI{611.1}{G} where the scattering length between fermions of opposite spin is $a=904 a_0$ \cite{zurn_precise_2013}, $1/k_F a = 7.4$, and the scattering length between the Feshbach molecules is $0.6 a$ \cite{ketterle_making_2008}. For non-degenerate clouds (a,b), the density distribution after the expansion is gaussian and therefore not condensed, however for degenerate clouds (c,d), the density is bimodal and therefore the entropy per fermion is below the critical value $(S/N k_B)_{c0} = 1.801$ for a non-interacting Bose gas in a harmonic trap \cite{pitaevskii_bose-einstein_2003}. The condensate fraction $N_0/N$ measured on the BEC side, which is independent of $\alpha$ and $\chi$, vs. the entropy per particle measured at unitarity (e) shows the critical point as a kink at the transition between where the condensate fraction is flat to where it monotonically increases as $S/N k_B$ decreases. The cloud is distorted due to the sweep of the magnetic trap frequency, so a finite condensate fraction is always fitted, and the critical entropy per fermion is set by the kink rather rather then the extrapolated $x$-intercept. Assuming the theoretical absorption cross section ($\alpha=1$), we find the measured critical entropy per particle $(S/N k_B)_c$ to be larger than $(S/N k_B)_{c0}$, indicating that the true entropy per particle is lower than our uncorrected thermometry indicates. We scale the cross section $\alpha$ to 2.15 such that $(S/N k_B)_c = (S/N k_B)_{c0}$ at the onset of condensation. We have verified that slower sweeps, sweeping further into the BEC regime with smaller scattering length, and longer time of flight do not change the measured condensate fraction. The reduced slope of $N_0/N$ vs. $S/Nk_B$ relative to the theoretical prediction may be due to residual weak repulsive interactions \cite{pitaevskii_bose-einstein_2003} or heating and three-body losses during this sweep. Both these effects also shift the transition to a lower $(S/N k_B)_c$, so the true $S/N k_B$ is likely slightly smaller than this calibration procedure indicates. The direct fit of $\alpha$ with the EoS is consistent with 2.15, but the systematic uncertainty is larger due to the low degeneracy of the clouds. We have also cross-checked this value of $\alpha$ calibrated at unitarity by measuring conductance quantisation to $1/h$ in the non-interacting gas, which relies on accurate thermometry to compute the compressibility \cite{krinner_observation_2014}.

The calibrated value $\alpha=2.15$ is similar to previously reported values \cite{ku_thermodynamics_2015, horikoshi_appropriate_2017} though the origin of this large correction factor is not completely clear. It is important to note that miscalibration of the imaging system's magnification and in $\chi$ is absorbed into $\alpha$, though we expect these corrections to be small: the 1\% (4\%) relative uncertainty in magnification ($\chi$) leads to 3\% (3\%) relative uncertainty in $S/N k_B$. $\chi/\alpha$ can be independently calibrated with the camera's magnified pixel size, quantum efficiency, gain, and the transmission of the imaging system \cite{horikoshi_appropriate_2017, hueck_calibrating_2017} though we find that this value is 50\% larger than the above calibration procedure predicts. It is therefore likely that there are still some non-ideal effects that are not explicitly accounted for in Eq.~\ref{eq:OD_from_images} but are approximately absorbed into $\chi$ and $\alpha$. Indeed, as in \cite{horikoshi_appropriate_2017}, we measure higher atom numbers at lower intensities, which would reduce the value of $\alpha$ required to match $(S/N k_B)_c = (S/N k_B)_{c0}$, and we observe the speckle pattern of the beam being partially visible in the OD at both low and high intensities. We can reasonably rule out density effects on the absorption cross section, e.g. from multiple scattering of photons \cite{veyron_quantitative_2022}, by verifying that the measured atom number is independent of the time of flight after release from the optical trap which significantly varies the density of the gas at fixed atom number. Imperfections of the beam's polarisation can increase $\alpha$, though this effect should be small as we measured with a polarisation analyser that the polarisation is within $1^\circ$ of the expected horizontal polarisation before and after the vacuum chamber. Even though we use short pulses, the Doppler shift can still be significant: Not only do the atoms acquire an average Doppler shift due to the absorbed photon recoil, they also develop an increasingly broad distribution due to the stochasticity in the direction of the re-emitted photons \cite{kazantsev_mechanical_1990}. We estimate that this can increase $\alpha$ by 10-20\%. For the short pulses we use, the atoms are not in the steady internal state that is assumed by Eq.~\ref{eq:OD_from_images}; we find by solving the Bloch equations for the 6 ground states and the 18 excited states in the $D_1$ and $D_2$ lines, we find that the number of scattered photons and therefore $\sigma$ is reduced by about 5\% relative to the steady state solution. The laser spectrum, both its lorentzian line and broad background typical of diode lasers, can also increase $\alpha$ by 5-10\%. Optical effects such as aberrations and refraction by the second spin state are also present but their effect on absorption imaging is rarely considered. A complete explanation of the deviation of $\alpha$ and $\chi$ from their expected values is beyond the scope of this study and we rely on the condensate fraction as an absolute calibration of $S/N k_B$. Moreover, none of our conclusions change qualitatively when not implementing this calibration procedure, only the value $\alpha_c$ becomes a factor of $\approx2$ larger and $I_\mathrm{exc}$ is $2.15$ times smaller.

\section{Fitting procedure}
\label{sec:fitting_procedure}

We fit the phenomenological model to each data set--the set of different transport times at fixed $\nu_x$--independently for both the first and second experiment. We do this by solving the initial value problem for $\Delta N(t)$ and $\Delta S(t)$
\begin{equation}\begin{split}
    \dv{\Delta N(t)}{t} &= -2 I_N[\Delta\mu(t), \Delta T(t)]
    = -2\qty{ I_\mathrm{exc} \tanh\bqty{\frac{\Delta\mu(t) + \alpha_c \Delta T(t)}{\sigma}} + G[\Delta\mu(t) + \alpha_c \Delta T(t)] } \\
    \dv{\Delta S(t)}{t} &= -2 I_S[\Delta\mu(t), \Delta T(t)]
    = -2\qty{\alpha_c I_N[\Delta\mu(t), \Delta T(t)] + G_T \Delta T(t)/T} \\
    \pmqty{\Delta\mu(t) \\ \Delta T(t)} &= \frac{2}{\kappa \ell_r} \pmqty{\ell_r + \alpha_r^2 & -\alpha_r \\ -\alpha_r & 1} \pmqty{\Delta N(t) \\ \Delta S(t)}
\end{split}\end{equation}
given the parameters $G$, $\alpha_c$, $I_\mathrm{exc}$, $\sigma$, $G_T$ along with the reservoir response functions $\kappa$, $\alpha_r$, $\ell_r$ and average temperature $T$ during transport and the initial values $\Delta N(0)$, $\Delta S(0)$. From these solutions, we also compute the total entropy as a function of time
\begin{equation}
    S(t) = S_\mathrm{eq} - \frac{\Delta N^2(t)}{2 T \kappa} - \frac{[\Delta S(t) - \alpha_r \Delta N(t)]^2}{2 T \ell_r \kappa}
\end{equation}
where $S_\mathrm{eq}$ is the equilibrium total entropy. From the data--the set of times $t_i$, relative particle imbalances $\Delta N_i/N_i$, relative entropy imbalances $\Delta S_i/N_i k_B$, and relative total entropies $S_i/N_i k_B$--we perform a least-squares fit by minimising the reduced chi-squared statistic
\begin{equation}
    \label{eq:redchi}
    \chi^2 = \frac{1}{\nu} \sum_i \pqty{\frac{\Delta N_i/N_i - \Delta N(t_i)/N}{\sigma_{\Delta N/N}}}^2
    + \pqty{\frac{\Delta S_i/N_i k_B - \Delta S(t_i)/N k_B}{\sigma_{\Delta S/N k_B}}}^2
    + \pqty{\frac{S_i/N_i k_B - S(t_i)/N k_B}{\sigma_{S/N k_B}}}^2.
\end{equation}
In other words, we simultaneously fit the particle imbalance, entropy imbalance, and total entropy. $\nu$ is the number of degrees of freedom: the number of data points minus number of fit parameters. For each quantity $x=\Delta N/N, \Delta S/N k_B, S/N k_B$, we use the same uncertainty for each data point defined as the average over each point's statistical uncertainty computed from 3-5 shots $\sigma_x = \sum_i \sigma_{x,i}/n_\mathrm{points}$ in order to improve the stability of the fits. We use the relative quantities normalised by the atom number (e.g. $\Delta N/N$ instead of $\Delta N$) to minimise the effect of $\sim\SI{10}{\percent}$ shot-to-shot fluctuations of the total atom number.

For both experiments, we set $G=0$ as allowing $G$ to be a fit parameter underconstrains the fit ($G$ is strongly correlated to $I_\mathrm{exc}/\sigma$) and enforcing $G=2 n_m/h$ increases $\chi^2$ by $\sim20\%$. We also fix $\kappa = \kappa^\mathrm{tr}(\mu^\mathrm{tr}, T^\mathrm{tr})$ to the value computed from the equation of state in the transport trap (see Sec.~\ref{subsec:effect_of_the_potential_energy_landscape}). For the first experiment, we fit $I_\mathrm{exc}$, $\sigma$, $\alpha_c$, $\Delta S(0)$, and $S_\mathrm{eq}$, and fix $G_T=0$ (the dynamics are dominated by the advective mode), $\alpha_r = \alpha_r^\mathrm{tr}(\mu^\mathrm{tr}, T^\mathrm{tr})$, and $\ell_r = \ell_r^\mathrm{tr}(\mu^\mathrm{tr}, T^\mathrm{tr})$. $\Delta S$ is offset by the value at long times where equilibrium is reached. This offset is a systematic bias in the thermometry due to drifts in the alignment.
For the second experiment, we fit $\sigma$, $\Delta S(0)$, $S_\mathrm{eq}$, $G_T$, $\alpha_r$, and $\ell_r$. As equilibrium is not reached at the longest time for $\nu_x>6$ kHz, we also fit an offset in $\Delta S$ to account for misalignment. Furthermore, we fix $\alpha_c=1.18k_B$ to the average fit result from the first experiment since only $\alpha_c-\alpha_r$ is important in the second experiment and we fix $I_\mathrm{exc}$ to the initial current $I_N(0)$ determined by a linear fit to all points in the initial response with $\Delta N/N \leq 0.07$. Without fixing $I_\mathrm{exc}$ in this way, it is strongly correlated to $\sigma$ and $G_T$ and yields similar values of $\chi^2$. Fixing $\alpha_r$ and $\ell_r$ to their theoretical values prevents the fit from reproducing the large $\Delta N$ that is induced by the initial entropy imbalance $\Delta S(0)$ in the second experiment.

\begin{figure}
    \centering
    \includegraphics{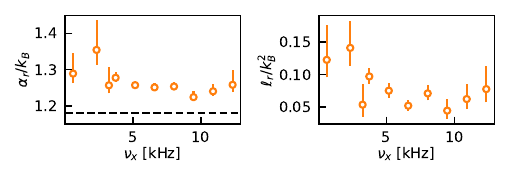}
    \caption{Fitted reservoir response coefficients for each configuration of the second experimentThe dashed horizontal line indicates the fixed $\alpha_c$ value in these fits.}
    \label{fig:reservoir_response_coefficients}
\end{figure}

With this method, the resulting $\chi^2 \approx 1$ for all data sets. The fitted values of the reservoir response coefficients from the second experiment are shown in Fig.~\ref{fig:reservoir_response_coefficients}. While $\alpha_r$ is relatively near the expected value for the anharmonic trap, $\ell_r$ is significantly smaller. There can be many reasons for these systematic biases, for example miscalibration of the absorption cross section and magnification, the contacts defined by the gate beam and wire forming smaller reservoirs contacted to the larger ones which develop their own effective thermodynamic response functions, or generally deviations of the true equation of state from the computed one. Despite this deviation from our expectations, this result does not change the overall form of the phenomenological model nor the central result of the large advective entropy current. We find that fixing $\alpha_r$ and $\ell_r$ in the first experiment yield good fits while they must be allowed to vary in the second experiment to fit the data. Allowing them to vary in the first experiment does not change the extracted parameters $I_\mathrm{exc}$ and $\alpha_c$. The average fitted nonlinearity from all the data sets is $\sigma = \SI{7.2(5)}{nK} \times k_B = \num{4.4(5)e-3}\Delta$.

We can derive an expression for the instantaneous slope of the path through state space $\Pi = \dv*{\Delta S}{\Delta N}$ by combining the expression for $I_S$ in Eq.~\ref{eq:currents_full_form} and the reservoir responses in Eq.~\ref{eq:reservoir_response}
\begin{equation}
    \Pi = \dv{\Delta S}{\Delta N} = \frac{\dv*{\Delta S}{t}}{\dv*{\Delta N}{t}} = \alpha_c + \frac{4 G_T (\alpha_r \Delta N - \Delta S)}{T \kappa \ell_r \dv*{\Delta N}{t}}
\end{equation}
In the limit where either the advective or diffusive mode dominates, the path is linear $\Delta S(t) \approx \Pi[\Delta N(t) - \Delta N_0]$ (Figs.~1d and 3f). When combined with the previous expression, this gives
\begin{equation}
    \Pi \approx \frac{4 G_T \alpha_r \Delta N + T \kappa \ell_r \alpha_c \dv*{\Delta N}{t}}{4 G_T (\Delta N - \Delta N_0) + T \kappa \ell_r \dv*{\Delta N}{t}}
\end{equation}
If the transport is fast as is the case for the superfluid-induced advective mode, we observe that $\Pi$ is independent of $\Delta N$, implying that the $\dv*{\Delta N}{t}$ terms dominate and the slope simplifies to
\begin{equation}
    \Pi_a \approx \alpha_c.
\end{equation}

We observe in the second experiment that after the advective mode has quickly relaxed and the diffusive mode is dominant, then the time evolution is exponential $\dv*{\Delta N}{t} \approx -\Delta N/\tau_d$, indicating that the response becomes linear in this regime $I_\mathrm{exc} \tanh[(\Delta\mu+\alpha_c\Delta T)/\sigma] \approx I_\mathrm{exc} (\Delta\mu+\alpha_c\Delta T)/\sigma$. Linearising tanh is equivalent to applying the linear model $\Xi = \Xi_n^a + \Xi_n^d$ with effective conductance $\tilde{G} = I_\mathrm{exc}/\sigma + G$ and Lorenz number $L = G_T / T \tilde{G}$
\begin{equation}
    \dv{t} \pmqty{\Delta N(t) \\ \Delta S(t)} \approx -2\tilde{G} \pmqty{1 & \alpha_c \\ \alpha_c & L + \alpha_c^2} \pmqty{\Delta \mu(t) \\ \Delta T(t)}
    = -\frac{4\tilde{G}}{\kappa \ell_r} \pmqty{1 & \alpha_c \\ \alpha_c & L + \alpha_c^2} \pmqty{\ell_r + \alpha_r^2 & -\alpha_r \\ -\alpha_r & 1} \pmqty{\Delta N(t) \\ \Delta S(t)}
    = -\vb{\Lambda} \pmqty{\Delta N(t) \\ \Delta S(t)}.
\end{equation}
In this limit, the path the system traces though state space is a line passing through equilibrium and can be written
\begin{equation}
    \pmqty{\Delta N(t) \\ \Delta S(t)} = \Delta N(0) e^{-t/\tau_d} \pmqty{1 \\ \Pi_d}
\end{equation}
so the exponential timescale $\tau_d$ and the slope of the path $\Pi_d$ are then given by the smaller eigenvalue and corresponding eigenvector of the matrix $\vb{\Lambda}$

\begin{align}\begin{split}
    \tau_d^{-1} &= \frac{2\tilde{G}}{\kappa\ell_r} \qty{L + \ell_r + (\alpha_c-\alpha_r)^2 - \sqrt{[L + \ell_r + (\alpha_c-\alpha_r)^2]^2 - 4L\ell_r}} \\
    &= \frac{4\tilde{G}}{\kappa \ell_r} [\ell_r + (\alpha_r - \alpha_c)(\alpha_r - \Pi_d)] \\
    \Pi_d &= \frac{L - \ell_r + \alpha_c^2 - \alpha_r^2 - \sqrt{[L + \ell_r + (\alpha_c-\alpha_r)^2]^2 - 4L\ell_r}}{2(\alpha_c-\alpha_r)} \\
    &= \frac{4\tilde{G}[\ell_r + \alpha_r(\alpha_r - \alpha_c)] + \kappa \ell_r \tau_d^{-1}}{4\tilde{G}(\alpha_r-\alpha_c)}.
\end{split}\end{align}
In the non-equilibrium steady state, $\tau_d^{-1} \rightarrow 0$ which implies that $[L + \ell_r + (\alpha_c-\alpha_r)^2]^2 \gg 4L\ell_r$, i.e., $L\rightarrow0$. Taylor expanding $\tau_d^{-1}$ and $\Pi_d$ to first order in $L$ yields

\begin{align}\begin{split}
    \tau_d^{-1} &\approx \frac{4\tilde{G} L}{\kappa[\ell_r + (\alpha_c-\alpha_r)^2]} \\
    \Pi_d &\approx \alpha_r + \frac{\ell_r}{\alpha_r - \alpha_c} \bqty{1 - \frac{L}{\ell_r + (\alpha_r-\alpha_c)^2}}.
\end{split}\end{align}
As first observed in \cite{husmann_breakdown_2018}, we see that $L$ decreases with increasing $\nu_x$ from Fig.~3 where the diffusive timescale $\tau_d$ becomes longer and from Fig.~4d where $\Pi_d$ increases.

\end{document}